\documentclass[useAMS,usedcolumn]{mn2e}
\usepackage{supertabular,epsf,rotate}
\newcommand{\f}{\phantom{2}}
\newcommand{\mc}{\multicolumn}

\newcommand{\ltsimeq}{\raisebox{-0.6ex}{$\,\stackrel 
        {\raisebox{-.2ex}{$\textstyle <$}}{\sim}\,$}}

\begin{document}

\title[Optical spectroscopy of radio galaxies in the 7CRS] {Optical
spectroscopy of radio galaxies in the 7C Redshift Survey}

\author[Willott et al.]{Chris J.\ Willott$^{1}$\footnotemark, Steve
Rawlings$^{1}$, Katherine M.\ Blundell$^{1}$, Mark Lacy$^{1,2,3}$, \and
Gary J. Hill$^{4}$ and Susan E. Scott$^{1,5}$ \\
$^{1}$Astrophysics, Department of Physics, Keble Road, Oxford, OX1
3RH, U.K. \\
$^{2}$Institute of Geophysics and Planetary Physics, L-413 Lawrence
Livermore National Laboratory, Livermore, CA 94550, USA \\
$^{3}$Department of Physics, University of California, 1 Shields
Avenue, Davis CA 95616, USA\\
$^{4}$McDonald Observatory, University of Texas at Austin, RLM
15.308, Austin, TX78712-1083, USA \\
$^{5}$Institute for Astronomy, Department of Physics and Astronomy,
The University of Edinburgh, Edinburgh EH9 3HJ}

\maketitle

\begin{abstract}

We present optical spectroscopy of all 49 radio galaxies in the 7C--I
and 7C--II regions of the 7C Redshift Survey (7CRS). The
low--frequency (151 MHz) selected 7CRS sample contains all sources
with flux-densities $S_{151} > 0.5$ Jy in three regions of the sky;
7C--I and 7C--II were chosen to overlap with the 5C6 and 5C7 surveys
respectively, and cover a total sky area of 0.013 sr.  The sample has
been completely identified and spectroscopy of the quasars and
broad-lined radio galaxies has been presented in Willott et
al. (1998). Only seven of the radio galaxies do not have redshift
determinations from the spectroscopy, giving a redshift completeness
for the sample of $> 90$\%. The median redshift of the 7CRS is 1.1.
We present a composite $0.2<z<0.8$ 7CRS radio galaxy spectrum and
investigate the strengths of the 4000 \AA\ breaks in these radio
galaxies. We find an anti-correlation between the 4000 \AA\ break
strength and emission line luminosity, indicating that departures from
old elliptical galaxy continuum shapes are most likely due to
non-stellar emission associated with the active nucleus.

\end{abstract}


\footnotetext{Email: cjw@astro.ox.ac.uk}

\section{Introduction}

Many attempts to deduce the nature of radio-loud AGN have depended
heavily on observations of 178 MHz selected 3C sources (e.g. Longair
1999). Studies of a revised 3C (3CRR) sample (Laing, Riley \& Longair
1983) have yielded valuable information on the nature and evolution of
the population of steep-spectrum radio sources. However the tight
correlation between radio luminosity and redshift in any single
flux-limited sample such as 3CRR make distinguishing between trends
with luminosity and redshift impossible. Most other completely
identified samples have been selected at high-frequencies (e.g. the
2.7 GHz 2 Jy sample of Wall \& Peacock 1985). High-frequency selected
samples contain different types of radio sources, such as a high
fraction of flat-spectrum quasars which have their jet axes close to
our line-of-sight. It is therefore difficult to compare the properties
of the low-frequency 3CRR sample with one selected at a much higher
frequency.

The first low-frequency selected sample fainter than 3CRR and
completely identified was the 6CE sample (Eales 1985; Rawlings, Eales
\& Lacy 2001). This sample contains every radio source in 0.1 sr of
sky with 151 MHz flux-density $2.0\leq S_{151}<3.93$ Jy. Hence it
contains sources with luminosities approximately a factor of 6 lower
than the 3CRR sample at any redshift. One example of the power of
investigating a range in radio luminosity at a given redshift is the
study which revealed that the $K$-band magnitudes of 6CE radio
galaxies are systematically fainter than those in the 3CRR sample at
$z \sim 1$ (Eales et al. 1997), indicating a relationship between the
host galaxy luminosity and radio luminosity in high-redshift radio
galaxies (Roche, Eales \& Rawlings 1998).

However, the fact that the 6CE sample is only a factor of 6 fainter
than 3CRR means that neither a large range in luminosity at a
particular redshift, nor a range in redshift at a constant luminosity
is achieved with the combination of the 3CRR and 6CE samples. Studies
such as deriving the luminosity function of radio sources are
therefore difficult with this small flux-density range. What is
required is another completely identified low-frequency sample with a
flux-density limit fainter than the 6CE sample. To this end we have
undertaken to obtain identifications and redshifts for radio sources
from three regions of the 7C catalogue -- the 7C Redshift Survey. In
this paper we present spectroscopic data on the radio galaxies in the
7C--I and 7C--II regions; data on all the quasars and broad-lined
radio galaxies (BLRGs) has been presented in Willott et al. (1998). Data on
the 7C--III region have been published in Lacy et al. (1999a), Lacy,
Bunker \& Ridgway (2000) and references therein.  

Using the 7CRS we have shown that the correlation between narrow
emission line and radio luminosities in radio galaxies and quasars
(Rawlings et al. 1989; Baum \& Heckman 1989) is real and not due to
correlations of these properties with redshift (Willott et al. 1999).
The 7CRS has also shown that the relative numbers of quasars and radio
galaxies is a strong function of both radio and narrow line
luminosity, with very few quasars at low luminosities (Willott et
al. 2000). Correlations between radio luminosity, redshift, linear
size and spectral index in the 3CRR, 6CE and 7CRS samples were
investigated with radio source evolution models to show that all high
redshift radio sources in these samples are observed a relatively
short time after the jet-triggering event (Blundell, Rawlings \&
Willott 1999). A new derivation of the radio luminosity function using
these samples has been presented by Willott et al. (2001).

In this paper we briefly discuss the selection of the sample (Section
2 -- a full description will be given in Blundell et al. in prep.) and
present optical spectroscopy of radio galaxies in the 7C--I and 7C--II
regions (Section 3). In Section 4 we give notes on individual sources.
In Section 5 we discuss the sample redshift distribution and the
stellar populations in the low-redshift ($z<0.8$) radio galaxies. The
seven sources without redshifts from optical spectroscopy (most of
which have very red optical to near-IR colours) are discussed in
Willott, Rawlings \& Blundell (2001; hereafter WRB).

\section{Sample selection} 

The seventh Cambridge (7C) survey was carried out with the Cambridge
Low Frequency Synthesis Telescope (CLFST; see McGilchrist et al. 1990
for details of the telescope) at a frequency of 151 MHz. The regions
7C--I and 7C--II overlap with fields 5C6 and 5C7, respectively, of the
5C survey (Pearson \& Kus 1978). An early attempt to identify these
sources was made by Rossitter (1987), but he was largely unsuccessful
due to the optical faintness of many of the sources.  These fields are
centred on $02^{\rm h} 14^{\rm m} 00^{\rm s},~ +32^{\circ} 00^{'}
00^{''}$ and $08^{\rm h} 17^{\rm m} 00^{\rm s},~ +27^{\circ} 00^{'}
00^{''}$ (epoch B1950.0), respectively.  The fields lie approximately
$30^{\circ}$ either side of the galactic plane, so there are few
problems of foreground obscuration and reddening of the optical
emission from the radio sources.

The completeness limit of the 7C survey in these fields is 0.12 Jy for
7C--I and 0.18 Jy for 7C--II (Blundell et al. in prep.). The 7C
Redshift Survey (7CRS) contains every source within these regions with
$S_{151}\ge 0.51$ Jy for 7C--I, $S_{151}\ge 0.48$ Jy for 7C--II and
$S_{151}\ge 0.5$ Jy for 7C--III. The slight difference between these
limits is due to a re-evaluation of the flux-density scales after much
of the follow-up observations had been made. The 7C--I region covers
0.0061 sr of sky and contains 37 objects, the 7C--II region covers
0.0069 sr and contains 39 objects. One source (3C 200) is common to
both the 3CRR sample and the 7CRS.

The resolution of the CLFST observations of the 7C fields is 70
$\times$ 70 cosec(Dec) arcsec$^2$. In order to determine the
structures of the sources, high-resolution observations were made with
the Very Large Array (VLA). Full details of the radio observations
will be given in Blundell et al. (in prep).

All members of the 7CRS (except the flat-spectrum quasar 5C7.230 and
3C200) have been imaged at $K$-band to provide an identification for
spectroscopy and to determine the near-IR morphology and magnitude.
Details of these observations and images are given in Willott et
al. (2002) where we discuss the $K-z$ relation and its dependence upon
radio luminosity. $K$ magnitudes for quasars and broad-lined radio
galaxies were also given in Willott et al. (1998). For almost all cases
a secure $K$-band identification has been found. In cases where
imaging gave more than one possible counterpart, long-slit
spectroscopy usually revealed the location of the true radio source
identification (see Section \ref{notes}).

To fix the astrometry of the $K$-band images, finding charts from
either the APM Catalogue (Irwin et al. 1994), or the HST Guide Star
Catalogue, were obtained for each radio source field. For most fields,
one or more stars detected on the $K$-band images were also on the
finding charts. For cases where three or more stars appear on both the
image and the chart, the {\footnotesize IRAF GASP} package was used to
determine the plate solution for the image.  For cases where only one
or two finding chart stars were detected on the $K$-band image, the
plate solution for another image with a good fit on the same observing
run was used along with the position of one of the detected stars. In
a few cases, no finding chart stars were within the image fields, but for
these objects wider field $R$-band images were available and the
astrometry could be achieved by first determining the positions of
fainter objects on the $R$-band images.

The VLA maps resolved the radio sources typically into core and lobe
components. Where a core is clearly identifiable, it is usual to find
the optical counterpart within an arcsecond of this position
(consistent with the residual error expected from the astrometric
process).  Where no core is visible, the counterparts are generally
found between the two lobes, often approximately equidistant between
the two. Where several objects were close to the expected counterpart
position, the radio and $K$-band images were overlaid. Subsequent
optical spectroscopy of candidates revealed the true identification
because of the strong emission lines present in most AGN.

\begin{table*}
\footnotesize
\begin{center}
\begin{tabular}{lccccccc}
\hline\hline
\mc{1}{l}{Name} &{Optical/Near-IR Position}&\mc{1}{c}{Telescope +}&\mc{1}{c}{Date}&\mc{1}{c}{Exposure}&\mc{1}{c}{Slit width}&\mc{1}{c}{Slit PA}&\mc{1}{c}{$z_{\rm spec}$} \\
\mc{1}{l}{ }  &\mc{1}{c}{(B1950.0)}&\mc{1}{c}{ Detector}&\mc{1}{c}{}& \mc{1}{c}{ time (s)}  & \mc{1}{c}{ (arcsec)}& \mc{1}{c}{ ($^{\circ}$)}\\
\hline\hline
 5C6.17     &  02 06 22.08~~  +34 14 25.2  &      ''   & 97Jan09 & 3600 &  2.5 & 170    & --    \\
 5C6.19     &  02 06 39.11~~  +33 40 09.8  &      ''   & 97Jan09 & 1546 &  2.5 & \f 95  & 0.799 \\
 5C6.24     &  02 07 20.18~~  +32 35 24.8  &      ''   & 95Jan29 & 1042 &  2.0 & \f 98  & 1.073 \\
 5C6.25     &  02 07 26.61~~  +33 56 38.3  &      ''   & 95Jan30 & 1254 &  3.1 & \f 16  & 0.706 \\
 5C6.29     &  02 08 09.00~~  +32 42 38.2  &      ''   & 95Jan31 & \f 600& 2.7 & \f 69  & 0.720 \\
 5C6.43     &  02 09 04.70~~  +33 48 14.3  &      ''   & 95Jan30 & 1800 &  3.1 & \f 16  & 0.775 \\
 5C6.62     &  02 10 29.69~~  +32 54 01.0  &      ''   & 95Jul28 & 1800 &  3.0 & \f 33  & --    \\
 5C6.63     &  02 10 29.85~~  +34 04 29.9  &      ''   & 95Jan31 & \f 994& 3.1 & \f 90  & 0.465 \\
 5C6.75     &  02 11 06.09~~  +30 12 15.2  &      ''   & 95Jan29 & 1456 &  3.1 & 145    & 0.775 \\
 5C6.83     &  02 11 11.17~~  +30 39 49.3  &      ''   & 97Jan09 & 2700 &  2.5 & 180    & --    \\
 5C6.78     &  02 11 17.94~~  +32 37 06.6  & McD+IGI   & 96Feb17 & 3600 &  2.0 & \f \f 0& 0.263 \\
 5C6.201    &  02 16 16.49~~  +34 05 56.1  & WHT+ISIS  & 98Dec21 & 1200 &  2.5 & 132    & 0.595 \\
 5C6.214    &  02 16 38.28~~  +34 09 25.6  &      ''   & 95Jan31 & 1800 &  2.2 & \f \f 0& 0.595 \\
 5C6.217    &  02 16 49.62~~  +33 34 31.5  &      ''   & 97Jan09 & 1800 &  2.5 & \f \f 0& 1.410 \\
 5C6.233    &  02 17 38.79~~  +29 38 35.6  &      ''   & 95Jan30 & 2700 &  4.0 & 350    & 0.560 \\
 5C6.239    &  02 18 10.60~~  +30 12 04.5  &      ''   & 95Jan31 & 1412 &  2.2 & \f 20  & 0.805 \\
 5C6.242    &  02 18 17.50~~  +31 03 24.8  &      ''   & 95Jan29 & 4380 &  2.7 & \f 25  & --    \\
 5C6.258    &  02 19 17.29~~  +33 39 49.1  &      ''   & 95Jan29 & 1600 &  1.8 & \f 77  & 0.752 \\
 5C6.267    &  02 20 07.00~~  +30 08 49.5  &      ''   & 97Jan09 & \f 300& 3.0 & \f 34  & 0.357 \\
 5C6.279    &  02 20 47.17~~  +34 01 36.4  &      ''   & 97Jan09 & \f 870& 3.0 & 359    & 0.473 \\
 7C0221+3417&  02 21 55.60~~  +34 17 04.1  &      ''   & 97Jan10 & 1800 &  2.5 & 146    & 0.852 \\
 5C6.292    &  02 23 54.99~~  +33 51 28.9  &      ''   & 94Jan09 & 1800 &  2.9 & 131    & 1.241 \\
 5C7.7      &  08 07 23.52~~  +26 59 44.5  & McD+IGI   & 96Feb17 & 1800 &  2.0 & \f \f 0& 0.435 \\
 5C7.8      &  08 07 59.52~~  +28 22 50.1  & WHT+ISIS  & 95Jan31 & 1216 &  2.9 & \f 37  & 0.673 \\
 5C7.9      &  08 08 18.14~~  +24 58 32.7  & McD+IGI   & 96Mar28 & 1200 &  2.0 & \f \f 0& 0.233 \\
 5C7.10     &  08 08 24.94~~  +26 27 14.8  & WHT+ISIS  & 95Jan29 & 1800 &  3.1 & \f 51  & 2.185 \\
 5C7.15     &  08 09 17.49~~  +26 39 17.3  &      ''   & 95Jan30 & 1800 &  2.0 & 270    & 2.433 \\
 5C7.23     &  08 10 33.38~~  +29 25 37.2  &      ''   & 95Jan29 & 1800 &  3.1 & 160    & 1.098 \\
 5C7.25     &  08 10 42.77~~  +28 00 46.6  &      ''   & 95Jan30 & 1645 &  2.0 & 278    & 0.671 \\
 5C7.47     &  08 12 38.78~~  +24 56 13.0  &      ''   & 00Feb10 & 3600 &  1.5 & 134    & --    \\
 5C7.57     &  08 13 29.64~~  +28 06 50.1  &      ''   & 97Apr07 & 1800 &  2.5 & 168    & 1.622 \\
 5C7.78     &  08 14 35.21~~  +29 14 31.1  &      ''   & 97Apr06 & \f 700& 2.5 & \f 17  & 1.151 \\
 5C7.79     &  08 14 39.24~~  +25 09 48.5  &      ''   & 97Apr06 & 1200 &  2.5 & \f 36  & 0.608 \\
 5C7.82     &  08 14 41.90~~  +29 31 11.5  &      ''   & 97Apr06 & \f 724& 2.5 & \f 90  & 0.918 \\
 5C7.106    &  08 15 51.26~~  +26 33 20.1  & McD+IGI   & 96Feb20 & 1200 &  2.0 & \f \f 0& 0.264 \\
 5C7.111    &  08 16 07.39~~  +26 24 11.1  & WHT+ISIS  & 97Jan09 & 1200 &  2.5 & \f 35  & 0.628 \\
 5C7.125    &  08 16 27.58~~  +25 28 48.2  &      ''   & 97Jan09 & 1543 &  2.5 & 119    & 0.801 \\
 5C7.145    &  08 17 21.19~~  +27 52 45.0  & McD+IGI   & 96Feb20 & 2400 &  2.0 & \f \f 0& 0.343 \\
 5C7.170    &  08 18 21.53~~  +25 28 36.9  &      ''   & 96Feb21 & 1200 &  2.0 & \f \f 0& 0.268 \\
 5C7.178    &  08 18 45.02~~  +29 32 19.4  & WHT+ISIS  & 97Jan09 & 1800 &  2.5 & \f 35  & 0.246 \\
 5C7.205    &  08 20 09.19~~  +24 40 17.1  &      ''   & 95Jan31 & \f 900& 3.4 & \f 71  & 0.710 \\
 5C7.208    &  08 20 18.74~~  +25 06 21.4  &      ''   & 95Jan31 & 3600 &  2.0 & \f 32  & --    \\
 5C7.223    &  08 21 05.57~~  +26 37 55.8  &      ''   & 95Jan29 & 1567 &  1.8 & \f 35  & 2.092 \\
 5C7.242    &  08 22 44.06~~  +24 50 11.0  &      ''   & 97Feb07 & 1200 &  2.5 & \f 60  & 0.992 \\
 5C7.245    &  08 22 55.99~~  +26 53 48.2  &      ''   & 95Jan31 & 3000 &  3.0 & 120    & --    \\
 5C7.269    &  08 25 39.48~~  +25 38 26.5  &      ''   & 95Jan31 & 1200 &  3.4 & 100    & 2.218 \\
 7C0825+2446&  08 25 45.26~~  +24 47 22.5  & McD+IGI   & 96Feb21 & 1200 &  2.0 & \f \f 0& 0.086 \\
 7C0825+2443&  08 25 56.62~~  +24 43 47.0  &      ''   & 96Feb21 & 2400 &  2.0 & \f \f 0& 0.243 \\
 5C7.271    &  08 26 00.89~~  +25 04 01.8  & WHT+ISIS  & 97Apr07 & 3600 &  2.5 & 105    & 2.224 \\
\hline\hline
\end{tabular}
\end{center}              
{\caption[Table of observations]{\label{tab:obstab} Log of
spectroscopic observations of radio galaxies in regions 7C--I and
7C--II of the 7CRS. The objects with `--' in the $z_{\rm spec}$ column
do not have spectroscopic redshifts from these observations. Their
likely redshifts are discussed in WRB.}}  \normalsize
\end{table*}

\section{Optical Spectroscopy} 
\label{optspec}

To attain the maximum scientific output from completely identified
radio samples, it is necessary to obtain redshifts for as many sources
as possible. The correlation between narrow emission line luminosity
and radio luminosity (e.g. Willott et al. 1999) means that radio
samples at fainter flux-density limits will also have lower flux
emission lines. This makes the task of achieving 100\% redshifts more
difficult for the 7CRS than, for example, the 3CRR sample. In spite of
this, over 90\% of the sources in 7C--I and 7C--II have spectroscopic
redshifts measured.

\subsection{Observations, reduction and analysis} 

Optical spectroscopy has been attempted for all members of the sample
except for 3C~200 ($z=0.458$; Spinrad et al. 1985). These observations
were made predominantly with the 4.2-m William Herschel Telescope
(WHT) on La Palma in the Canary Islands. A few of the brighter sources
have optical spectra obtained at the $107^{\prime \prime}$ telescope
at the McDonald Observatory, operated by the University of Texas at
Austin. A full list of the spectroscopic observations of the radio
galaxies is given in Table \ref{tab:obstab}. Conditions were
photometric for all observations except that of 5C6.292, where clouds
affected the transparency.

All the WHT spectra were obtained using the ISIS double-beam
spectrograph. This instrument uses a dichroic to separate the red and
blue light into different beams which are detected on separate
charge-coupled devices (CCDs). At that time, each TEK CCD comprised of
$1124^{2}$ pixels which were windowed down in the spatial direction to
$400$ pixels to reduce the readout times. Two 158 lines mm$^{-1}$
gratings were used to provide broad spectral coverage from 3000 \AA\
-- 5800 \AA\ on the blue arm and 5400 \AA\ -- 8400 \AA\ on the red
arm. The pixel scale in the dispersion direction was $\approx 2.5$
\AA\ pix$^{-1}$ and the resolution $\approx 10$ \AA. The scale along
the spatial direction was 0.358 arcsec pix$^{-1}$ giving a total slit
length on the sky of 143 arcsec. Slit widths ranged from 1 to 4
arcsec, depending upon the seeing and the optical/near-IR spatial
extent of the object. Position angles of slits on the sky were set to
include several objects if the identification was uncertain or a close
companion was to be observed simultaneously. In other cases the slit
was usually oriented along the radio axis to search for extended line
emission aligned with the radio axis.

\begin{figure*}
\epsfxsize=0.94\textwidth
\vspace{0.8cm}
\hspace{0.8cm}
\begin{centering}
\epsfbox{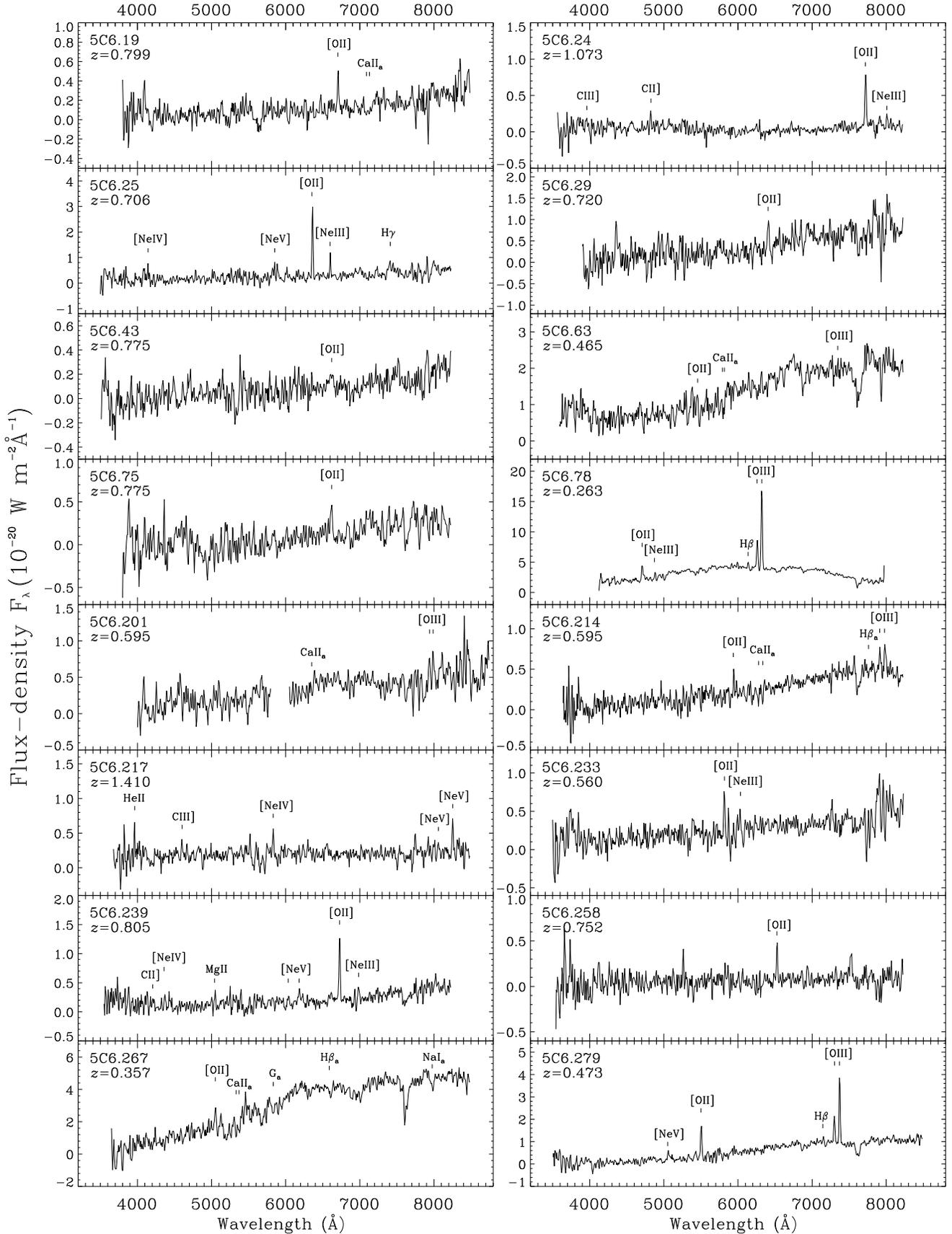} 
\end{centering}
\vspace{0.6cm}
{\caption[junk]{\label{fig:rgspec} Optical spectra of radio galaxies in
the 7C sample.}}
\end{figure*}	

\addtocounter{figure}{-1}

\begin{figure*}
\epsfxsize=0.94\textwidth
\vspace{0.8cm}
\hspace{0.8cm}
\begin{centering}
\epsfbox{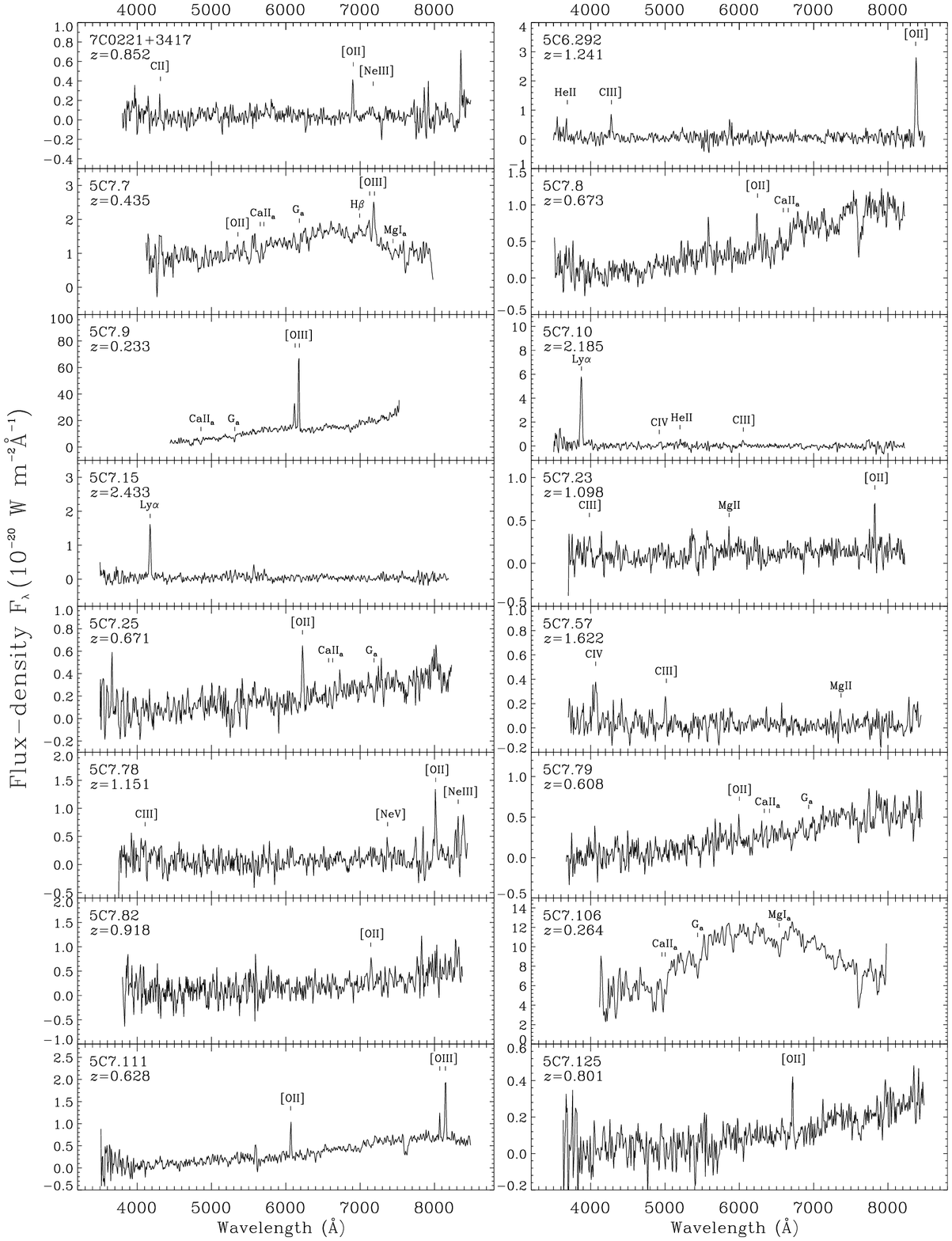} 
\end{centering}
\vspace{0.6cm}
{\caption[junk]{continued}}
\end{figure*}	

\addtocounter{figure}{-1}

\begin{figure*}
\epsfxsize=0.94\textwidth
\vspace{0.8cm}
\hspace{0.8cm}
\begin{centering}
\epsfbox{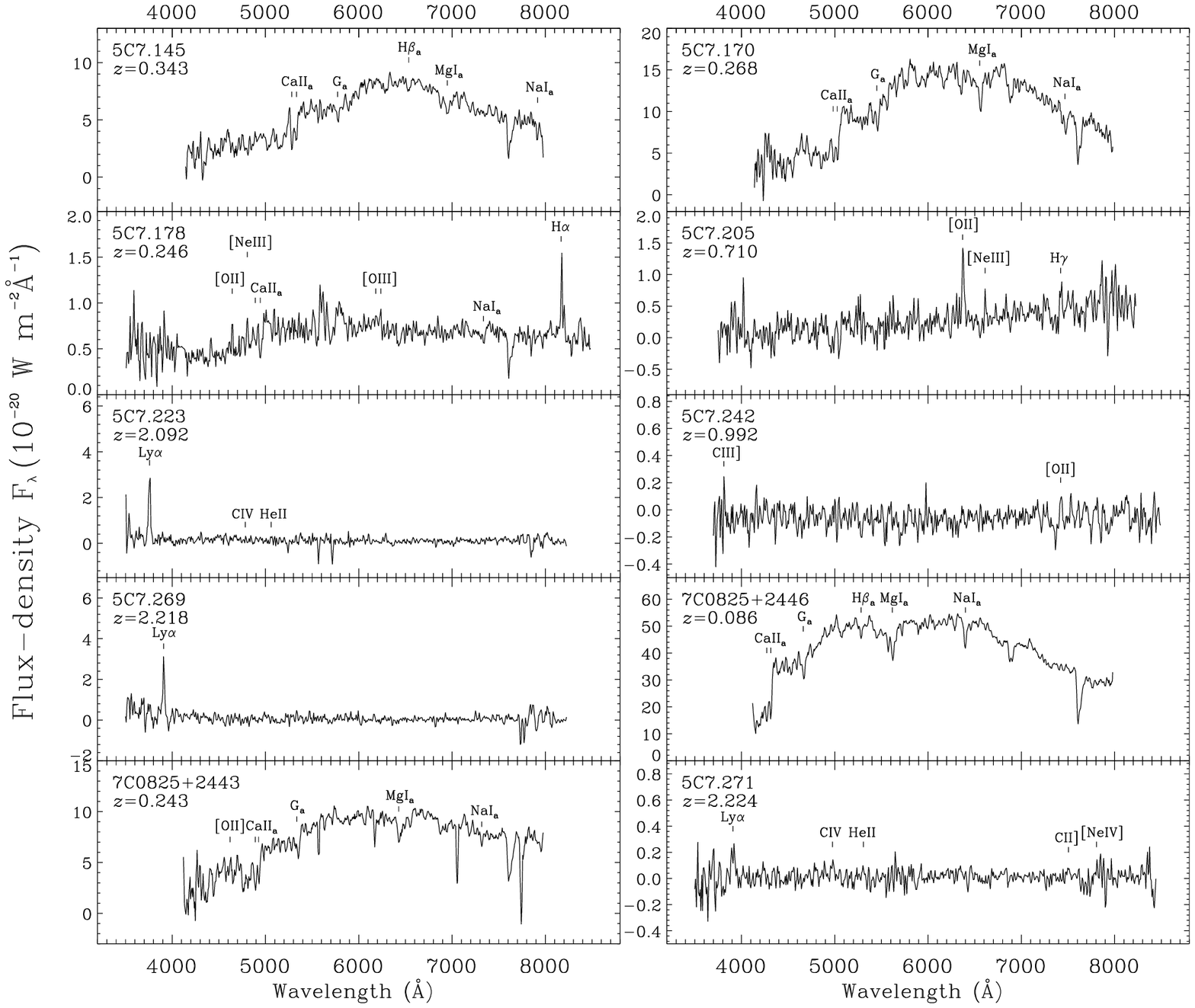} 
\end{centering}
\vspace{-7.6cm}
{\caption[junk]{continued}}
\end{figure*}	

The spectra were reduced using standard {\footnotesize IRAF}
procedures as described in Willott et al. (1998). In general, the
telescope pointing was very accurate and the object appeared in the
expected row ($\pm 2$ rows). In some ambiguous cases, other objects
along the slit were identified on the $K$-band images and the
astrometry of the spectra could be determined. One-dimensional spectra
of candidate objects were extracted from the two-dimensional
spectra. Two extractions were made for each source: a full-width at
half-maximum (FWHM) aperture for a good signal-to-noise ratio (snr)
and a full-width at zero-intensity (FWZI) aperture for flux
measurements. Atmospheric absorption features have not been corrected
for, since the residuals from such processes can often appear as real
features. Cosmic rays were identified in the 2-D spectra and edited
out of the 1-D spectra. Finally, the red and blue spectra (in the case
of ISIS observations) were joined together, averaging over 50 \AA\
where the red and blue sensitivity functions crossed.

The final FWHM-aperture spectra of all radio galaxies which allowed
redshift determinations are shown in Figure \ref{fig:rgspec}. Six
sources (5C6.17, 5C6.62, 5C6.83, 5C6.242, 5C7.208 and 5C7.245) did not
show any reliable continuum or emission lines in their spectra.
However, the pointing positions of the attempted spectra of these
sources are secure, since they have all been identified at
$K$-band. In addition, the spectrum of 5C7.47 has weak patchy
continuum which does not allow a redshift determination. For these
seven objects, we have obtained optical and near-IR imaging in order
to constrain their redshifts by fitting model galaxy spectra. We have
also obtained some near-IR spectroscopy to search for emission
lines. These observations and the resulting redshift constraints are
discussed fully in WRB. For all the other radio galaxies, emission
and/or absorption lines were identified which allowed a secure
redshift determination.

A gaussian fit was made to the lines on the FWHM aperture spectra to
determine the line centres and FWHMs. The redshift of each object was
determined from a weighted mean of emission and/or absorption line
centres. Several of the radio galaxies in the redshift range
$0.6<z<1.2$ show only one emission line which we identify as
redshifted [OII] $\lambda3727$ \AA. Continuum bluewards of the
emission line prevents it being Ly$\alpha$. Also, the continuum near
the emission line is rising with increasing wavelength and the stellar
4000 \AA\ break is observed, confirming the redshift. These sources
are described individually in Section \ref{notes}.

Emission line fluxes were measured with IRAF on the FWZI spectra, by
fitting a linear continuum and integrating the flux above this
continuum. The observed equivalent widths were also measured in this
way. The main source of error here is in accurately determining the
level of the underlying continuum. The errors quoted therefore
represent the typical uncertainty in the continuum level. Appendix A
contains the emission and absorption line data measured from all the
spectra.

\section{Notes on individual sources} 
\label{notes} 

For radio galaxies which do not have two or more reliable emission or
absorption lines in their optical spectra, we now discuss the evidence
for the redshifts quoted. In all cases unless stated, the $K$-band
magnitudes are consistent within 2$\sigma$ with the derived redshifts
according to the $K-z$ relation for 3CRR, 6CE and 7CRS radio galaxies of
Willott et al. (2002).\\

{\bf 5C6.19} has one strong emission line identified as [OII] $\lambda
3727$ and a rise in continuum redward of 7200 \AA\ consistent with a
4000 \AA\ break at the same redshift. The Ca H and K absorption
features are also marginally detected in the spectrum at the correct
redshift.\\

{\bf 5C6.29} has one strong emission line which we identify as [OII]
$\lambda 3727$. The rise in continuum redward of 7000 \AA\ is
consistent with a 4000 \AA\ break at the same redshift.\\

{\bf 5C6.43} has a probable [OII] $\lambda 3727$ line and a rise in
continuum redward of 7000 \AA\ consistent with a 4000 \AA\ break at
the same redshift.\\

{\bf 5C6.63} shows only weak, marginal emission and absorption lines,
but a strong continuum with a large 4000 \AA\ break at 5800\AA.\\

{\bf 5C6.75} has one reliable emission line identified as [OII]
$\lambda 3727$. Scattered light from a nearby star has been removed
fairly effectively from the spectrum, however this has led to an
increase in the noise level. There is an increase in continuum
strength redward of the emission line, consistent with its
identification as [OII]. \\

{\bf 5C6.201} is a large double-lobed radio source with a separation
between the hotspots of 90 arcsec. There is no radio core detected in
our current radio map. There is a $K=17.4$ galaxy approximately
equidistant between the hot spots which we believe to be the
identification. Although no definite emission or absorption features
are visible in the spectrum of this galaxy, there is probably a 4000
\AA\ break at $\approx 6350$ \AA\ and possibly [OIII] lines in the
noisy sky emission at 8000 \AA\ giving a redshift of $z=0.595$. At
this redshift the [OII] $\lambda 3727$ is in a region of high noise
due to the dichroic. This object has very weak emission lines as
expected for some low--redshift sources (c.f. 5C6.63) from the
correlation between narrow line and radio luminosities (Willott et
al. 1999). Note that the $K$-magnitude of this galaxy is about 1
magnitude fainter than the mean in the $K-z$ relation, but this is
within the 2$\sigma$ scatter about the relation, so does not rule out
this galaxy hosting the radio source. A close companion galaxy (7
arcsec to the north-west) could potentially be the host of the radio
source. This galaxy has a very similar $K$-magnitude and optical
spectrum and is probably physically associated with the other
galaxy. Therefore we are fairly confident of the redshift of the radio
source. We are planning further radio observations to determine which
of these two galaxies is the actual radio source host and then deeper
spectroscopic observations. The radio source 5C6.214 is only 5
arcminutes from 5C6.201 and has an identical redshift. Thus they are a
pair of powerful radio sources separated by only $\sim 2$ Mpc
($H_0=50~ {\rm km~s^{-1}Mpc^{-1}}$).\\

{\bf 5C6.258} shows one strong emission line. We identify this as [OII]
$\lambda 3727$ on the basis of the $K$ magnitude of 17.8 making this
line unlikely to be Ly$\alpha$ at $z=4.4$.\\

{\bf 5C7.15} has only one definite emission line. On the basis of a
lack of continuum and consequently high lower limit to the equivalent
width of the emission line and its faintness at $K$-band, this line is
identified as Ly$\alpha$ at $z=2.433$.\\

{\bf 5C7.82} has one strong emission line identified as [OII] $\lambda
3727$ and a rise in continuum redward of 7700 \AA\ consistent with a
4000 \AA\ break at the same redshift.\\

{\bf 5C7.125} has one strong emission line identified as [OII]
$\lambda 3727$ and a rise in continuum redward of this line consistent
with a 4000 \AA\ break at the same redshift.\\

{\bf 5C7.178} shows many narrow emission lines at $z=0.246$, as well
as a couple of stellar absorption features and a prominent 4000 \AA\
break. However, the unusual thing about this object is that it is
several magnitudes fainter than the mean $K-z$ relation and the only
object in the 7C sample which is a serious $K-z$ outlier (Willott et
al. 2002). The strength of high ionisation lines such as [Ne III]
compared to the Balmer lines indicates that the ionisation mechanism
for this source is probably photoionisation by an active nucleus and
not a starburst. The [OII] line strength is abnormally weak compared
to other 7C galaxies at this redshift. These two facts suggests the
radio luminosity of 5C7.178 is unusually high.\\

{\bf 5C7.242} has one definite emission line at 7427 \AA\ and a
low-significance line at 3814 \AA. These two lines are consistent
with a redshift of $z=0.992$. There is little continuum evident in the
spectrum and hence no clear 4000 \AA\ break is seen.\\

{\bf 5C7.269} has only one definite emission line in our spectrum. On
the basis of a lack of continuum and consequently high lower limit to
the equivalent width of the emission line and its faintness at
$K$-band, we would expect this line to be identified as Ly$\alpha$ at
$z=2.218$. We note that Eales \& Rawlings (1996) presented an optical
spectrum of this radio source which shows Ly$\alpha$ and Si IV
$\lambda1394$ at $z=2.218$ and a near-infrared spectrum showing a
marginal H$\alpha$ line at a consistent redshift. Simpson et al. (in
prep.) have obtained a deep near-infrared spectrum which confirms the
presence of [OII] and [OIII] emission lines at this redshift.

\begin{figure}
\epsfxsize=0.47\textwidth
\begin{centering}
\epsfbox{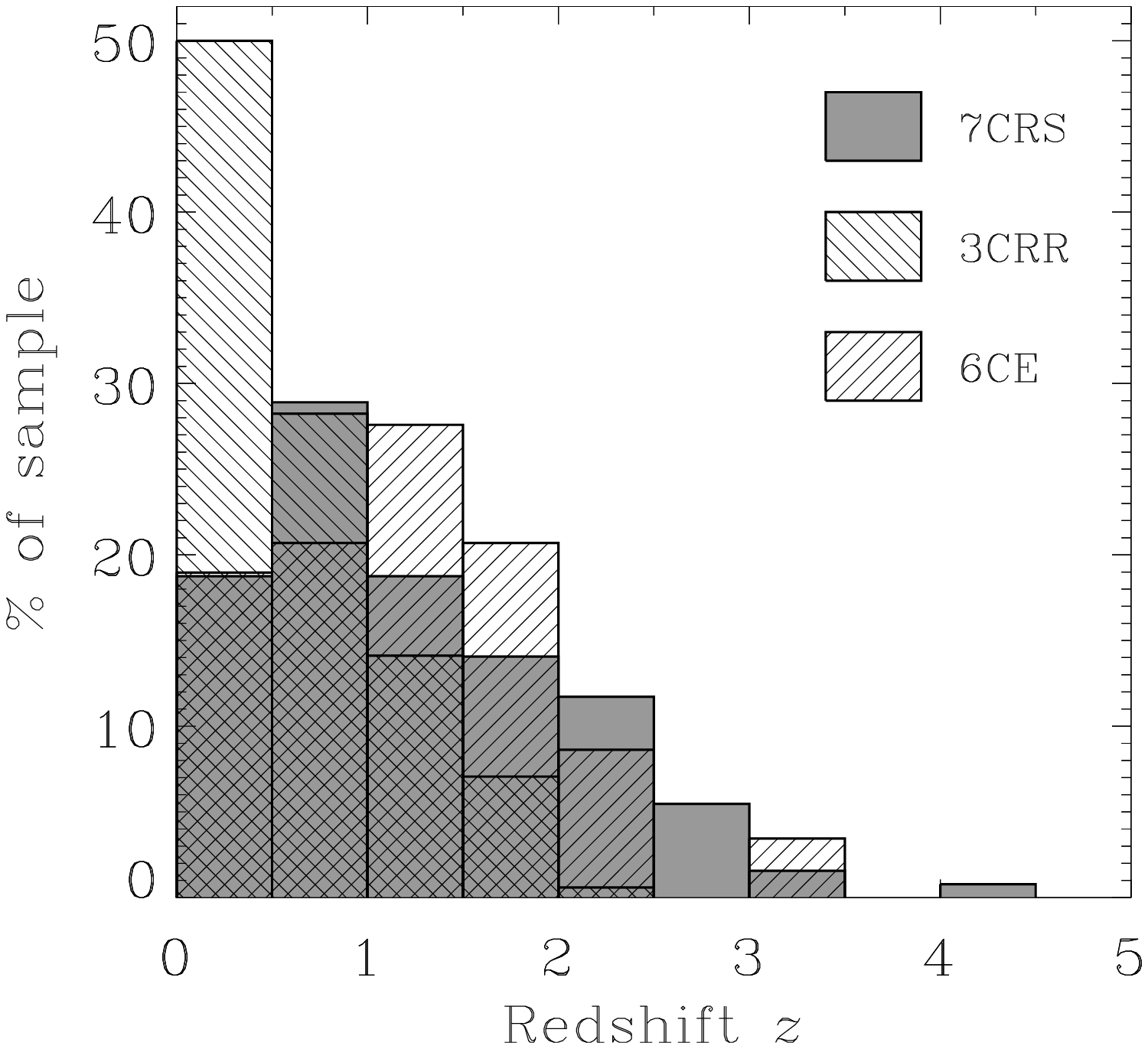} 
\end{centering}
{\caption[junk]{\label{fig:nz} Histogram of the redshift distributions
of sources in the 7CRS (regions I, II \& III), 3CRR and 6CE samples.
}}
\end{figure}

\section{Discussion}

\subsection{Redshift distribution}

In Fig. \ref{fig:nz} we show the redshift distribution of the 7CRS
(including regions I, II and III). The distribution peaks at $z \sim1$
and the median redshift is $z=1.1$. The redshift distribution is quite
different to that of the 3CRR sample, which peaks at low redshift
(median $z=0.5$). The 6CE sample has a redshift distribution very
similar to that of the 7CRS and has the same median redshift. A full
derivation of the radio luminosity function using these three samples
is given in Willott et al. (2001).

\subsection{Stellar content of 7CRS galaxies}

Most of the low redshift ($z<1$) 7CRS radio galaxy spectra show a
continuum dominated by stellar light, with absorption features visible
in many cases. Therefore we can use these spectra to investigate the
stellar content of the radio galaxies and in particular whether they
show evidence for recent starburst activity.  The snr of individual
spectra do not permit detailed fitting by stellar population
models. Instead, we investigate the size of the $4000$ \AA\ break in
each galaxy to estimate the amount of recent star-formation or
non-stellar UV component. In addition we can combine the radio galaxy
spectra to obtain a high snr composite and use this to compare with
stellar population models.

\begin{figure}
\epsfxsize=0.37 \textwidth
\hspace{0.9cm}
\begin{centering}
\epsfbox{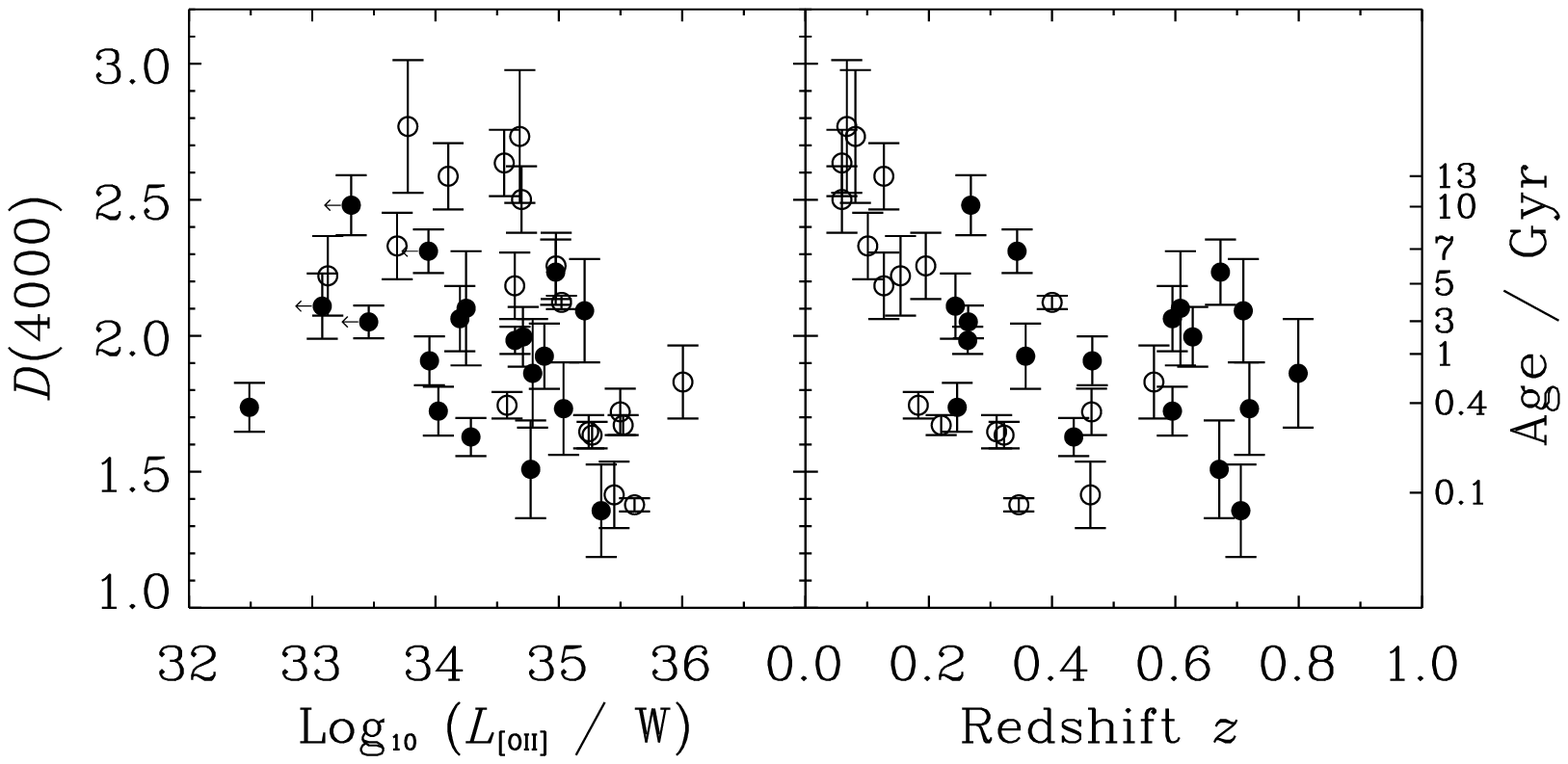} 
\end{centering}
\vspace{0.6cm} 

{\caption[junk]{\label{fig:d4000} The size of the $4000$ \AA\ break
$D(4000)$ against log [OII] emission line luminosity $L_{\rm [OII]}$
(left) and redshift $z$ (right) for all radio galaxies in 7C-I and
7C-II with $0.2<z<0.8$ (filled circles). Also plotted are radio
galaxies from the 2 Jy sample from Tadhunter et al. (2002) and 3CR
sample from Wills et al. (2002) (open circles). The right-hand y-axis
shows how $D(4000)$ depends upon the age of a single stellar
population. In both plots, there is an anti-correlation present at the
3$\sigma$ level. Emission line luminosities have been calculated
assuming $H_0=50~ {\rm km~s^{-1}Mpc^{-1}}$, $\Omega_{\rm M} =1$,
$\Omega_ \Lambda=0$. }} \end{figure}

The $4000$ \AA\ break feature is a sensitive indicator of the age of
the dominant stellar population, since a large break indicates few
massive, young stars in the galaxy (Bruzual 1983). The size of the
$4000$ \AA\ break $D(4000)$ was measured for all the radio galaxies at
$0.2 < z < 0.8$ in 7C--I and 7C--II. We used slightly different
wavelength bins compared to the definition of Bruzual in order to
avoid emission line contamination and to enable a consistent
comparison with the results of Tadhunter et al. (2002). We define
$D(4000)= \int_{4150}^{4250} F_\nu d\lambda~ / \int_{3750}^{3850}
F_\nu d\lambda$. Errors were estimated by adding in quadrature the
standard errors on the fluxes in each bin. The resulting values of
$D(4000)$ are given in Table A1.

\begin{figure*}
\epsfxsize=0.94\textwidth
\vspace{-0.4cm}
\begin{centering}
\epsfbox{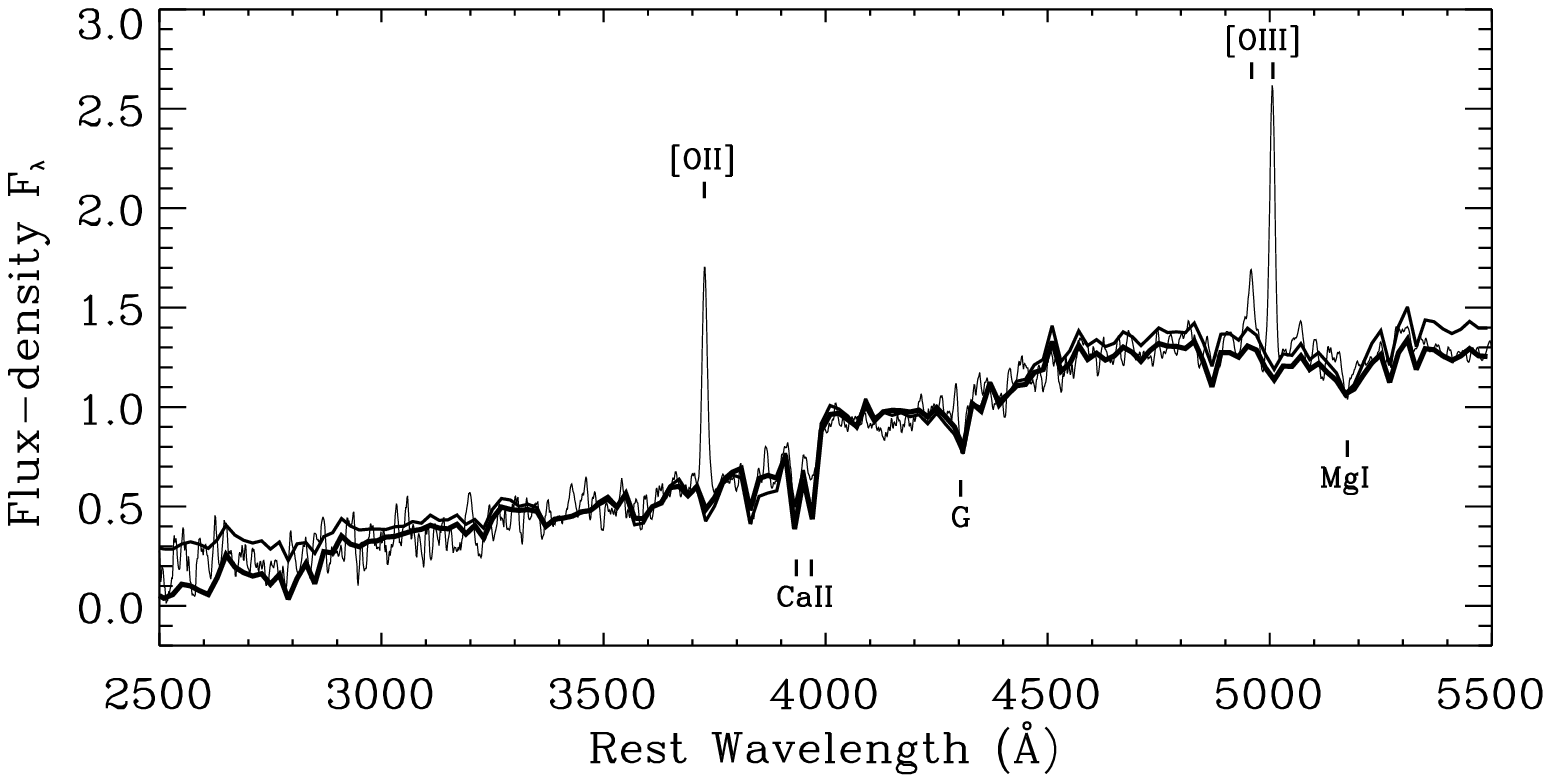} 
\end{centering}
\vspace{-0.3cm} 

{\caption[junk]{\label{fig:compgal} Composite 7CRS radio galaxy
spectrum created from all 7C--I/7C--II galaxies in the redshift range
$0.2<z<0.8$ (thin line). Also shown are a Bruzual \& Charlot (2002)
model stellar population which formed in an instantaneous burst 2.0
Gyr ago (thick line) and a two component model consisting of old (10
Gyr) and young (0.3 Gyr) components (medium line). Prominent emission
lines in the composite spectrum are labelled above the spectrum and
absorption lines labelled below. Note that the Bruzual \& Charlot
models have much lower spectral resolution than our data and therefore
the depths of absorption lines in the data and the model cannot be
compared. }}

\end{figure*}

In Fig. \ref{fig:d4000} we plot $D(4000)$ against narrow [OII]
emission line luminosity $L_{\rm [OII]}$ and redshift $z$ for all the
radio galaxies where $D(4000)$ could be reliably measured (error $<
0.25$). On the right-hand y-axis we show how $D(4000)$ corresponds to
the ages of instantaneous burst, single population Bruzual \& Charlot
(2001) models. The range of $D(4000)$ for the 7CRS galaxies
corresponds to ages of $0.1 - 10$ Gyr for a single stellar
population. There is marginal evidence for anti-correlations between
both $D(4000) - L_{\rm [OII]}$ and $D(4000)- z$ in the 7CRS, but
neither of these correlations are statistically significant at the 2
$\sigma$ level. Given that within a single flux-limited sample radio
luminosity and redshift are tightly correlated and $L_{\rm [OII]}$ is
correlated with radio luminosity, there is also an apparent
correlation between $L_{\rm [OII]}$ and $z$. In order to investigate
if $D(4000)$ is redshift or emission line luminosity dependent, we
have included data from two brighter samples of low-redshift radio
galaxies. Tadhunter et al. (2002) measures $D(4000)$ for radio
galaxies with $0.1<z<0.7$ from the 2 Jy steep-spectrum radio galaxy
sample (Tadhunter et al. 1998) and Wills et al. (2002) measure
$D(4000)$ for $z<0.2$ 3CR galaxies. We exclude the broad-line radio
galaxies from these samples, since they have systematically lower
values of $D(4000)$ than the narrow line galaxies, indicating the
presence of a strong non-stellar UV component. With the inclusion of
the 2 Jy and 3CR radio galaxies, there are clearly anti-correlations
between $D(4000)$ and $L_{\rm [OII]}$ and $D(4000)$ and $z$ (both
significant at the 3$\sigma$ level). The correlation with redshift is
due largely to the fact that the $z<0.2$ 3CR galaxies all have large
4000 \AA\ breaks. At low redshifts and/or [OII] luminosities, most
galaxies have large breaks ($D(4000)>2$) consistent with those found
for inactive ellipticals at similar redshifts (Hamilton 1985). At
higher luminosities, there is a greater spread in $D(4000)$ with many
sources having smaller breaks. We will return to the interpretation of
these results after examining the properties of the composite radio
galaxy spectrum.

Composite optical spectra of objects spanning a wide range of
redshifts have limited applications, since different objects will be
contributing at different parts of the rest-frame spectrum. Since the
high redshift galaxies in the 7CRS have only weak continua in their
optical spectra, we choose to include only sources in the redshift
range $0.2 < z < 0.8$. To construct the composite spectrum the optical
spectra of all the radio galaxies in this redshift range in the 7C--I
and 7C--II regions (totalling 25 galaxies) were trimmed to exclude
noisy regions at the ends and then shifted to the rest-frame using the
redshifts in Table \ref{tab:obstab}. All spectra were then normalised
by the integrated flux over the range $4000 - 4500$ \AA. This region
was chosen because it is in the rest-frame spectra of all the included
objects, it is devoid of bright emission lines and also lies above the
$4000$ \AA\ break. The spectra were then rebinned, merged and smoothed
with a 10 \AA\ boxcar filter to form the composite which is shown in
Fig.  \ref{fig:compgal}. Note that not all objects contribute across
this whole wavelength range and only the region $3650 - 4560$ \AA\
includes all 25 galaxies. However, the composite is reasonably
complete at $3000 - 5000$ \AA\ with 17 galaxy spectra going as low as
$3000$ \AA\ and 16 as high as $5000$ \AA. The median radio and
emission line luminosities of the radio galaxies comprising the
composite are $\log_{10} (L_{151}$~/~WHz$^{-1}$sr$^{-1}) =25.9$ and
$\log_{10} (L_{\rm [OII]}$~/~W$)=34.6$

The composite 7CRS radio galaxy spectrum shows strong narrow emission
lines of [OII] and [OIII]. The relative strengths of these lines in
the composite is not indicative of the ratios of these lines in 7CRS
radio galaxies, due to the facts that only the lower redshift (and
hence lower luminosity) sources have [OIII] in their optical spectra
and there is a strong correlation between the narrow line and radio
luminosities (Willott et al. 1999). However, because the contributing
spectra were normalised by their continua, the shape of the continuum
and the strength of the absorption lines of the composite are not
strongly affected by this problem (and certainly not in the $3650 -
4560$ region which contains all the galaxy spectra). The spectrum in
Fig. \ref{fig:compgal} has $D(4000)=2.0$.

We have compared the composite galaxy spectrum to a set of Bruzual \&
Charlot (2002) instantaneous burst stellar population models with
various ages ranging from 0.01 to 20 Gyr. All the models have a
Salpeter IMF, solar metallicity and are normalized over the range
$4000 - 4500$ \AA.  A least-squares fit of the models to the composite
spectrum (wavelength range 3000 - 5000 \AA\ only, with the emission
lines subtracted and rebinned to the model resolution) gives a best
fit for an age of 2.0 Gyr . This model fits the composite galaxy
spectrum remarkably well with reduced $\chi ^2 = 1.9$ -- this model is
also shown in Fig.  \ref{fig:compgal}.  This is particularly
surprising given that the composite spectrum is a combination of 25
galaxies with different star-formation histories and quite a range of
$D(4000)$. The composite radio galaxy spectrum can be equally well-fit
by a two population model of a dominant old (10 Gyr) population and a
young (0.3 Gyr) component with only 1.5\% of the total stellar mass
(Fig.  \ref{fig:compgal}).  A model comprising an old (10 Gyr) stellar
population and a power law can also be fit to the composite, resulting
in a slightly poorer fit (reduced $\chi ^2 = 3.0$). Models involving
significant young components with ages $\leq 0.1$ Gyr do not fit the
composite well showing that most of the radio galaxies have not
undergone recent starbursts involving 0.5\% or more of the stellar
mass (unless these starbursts are obscured by dust). This is
consistent with the fact that the $R-K$ colours of most $z<1$ 7CRS
radio galaxies are as expected for evolved galaxies which formed at
high redshifts (WRB).

From the 4000 \AA\ break analysis and the composite spectrum, we find
that the 7CRS radio galaxies have a range of continuum shapes which
can be accounted for by a dominant old stellar population and a
blue/UV component of varying strength. This blue component could be
due either to recent star-formation or non-stellar emission associated
with the active nucleus. At higher redshifts and luminosities, it is
well-known that there is a strong blue component, often aligned with
the radio jet axis (McCarthy et al. 1987). No single mechanism for
this alignment effect fits the data for all sources and scattered
quasar light, nebular continuum and jet-induced star-formation all
play a role (Best et al. 1998). Observations of the 2 Jy sample at
$0.15<z<0.7$ show that the UV excesses in these relatively
radio-luminous galaxies are due to a mixture of non-stellar emission
such as nebular continuum and direct AGN light and young stellar
populations (Tadhunter et al. 2002). The anti-correlation between
$D(4000)$ and $L_{\rm [OII]}$ for our combined sample shows that the
strength of the blue component decreases with emission-line
luminosity.  This therefore suggests that the strength of the blue
component is more closely linked to the active nucleus than to recent
star-formation. WRB found a similar correlation at higher redshifts
such that the reddest radio galaxies at $1<z<2$ in 7C--I and 7C--II
also had the weakest emission lines. Lacy et al. (1999b) showed that
for $0.5<z<0.8$ 7CRS radio galaxies there is a measurable alignment
effect, albeit weaker than that in more luminous radio
galaxies. Therefore, we conclude that non-stellar emission is likely
to be a more common cause for the blue excess in some 7CRS radio
galaxies than recent star-formation.

\section{Summary}

We have presented the results of optical spectroscopy of the 7C--I and
7C--II regions of the 7C Redshift Survey. All but 7 of the 76 radio
sources in the sample have spectroscopic redshifts. These seven
objects have redshifts constrained by optical/near-IR imaging and
near-IR spectroscopy (WRB). The high redshift completeness of this
sample makes it ideal for combining with the brighter, low-frequency
selected 3CRR and 6CE samples to study the nature and evolution of
radio sources and their host galaxies. At $z \ltsimeq 0.8$ 7CRS radio
galaxies typically have optical continua dominated by evolved stellar
populations. The UV excesses in some radio galaxies are most likely
due to to non-stellar AGN-related emission, rather than recent
star-formation. Early results using these data can be found in
Blundell, Rawlings \& Willott (1999) and Willott et al.
(1999,2000,2001). Summary tables of basic data on the 7C--I and 7C--II
samples can be found in Willott et al. (2002).

\section*{Acknowledgements}

We would like to thank Steve Eales, Julia Riley and David Rossitter
for important contributions to the 7C Redshift Survey. Thanks to Clive
Tadhunter and Karen Wills for providing us with data ahead of
publication and to Margrethe Wold for help with the McDonald Telescope
observations. We thank the referee Dr E.M. Sadler for a useful
referees report. Thanks to the staff of the ING for their excellent
support during the spectroscopic observations. The William Herschel
Telescope is operated on the island of La Palma by the Isaac Newton
Group in the Spanish Observatorio del Roque de los Muchachos of the
Instituto de Astrofisica de Canarias. Thanks to the staff of McDonald
Observatory for their support of part of these observations. This
research has made use of the NASA/IPAC Extra-galactic Database, which
is operated by the Jet Propulsion Laboratory, Caltech, under contract
with the National Aeronautics and Space Administration. CJW thanks
PPARC for support. GJH acknowledges support from the Texas Advanced
Research Program under Grant No. 009658-0710-1999.

\appendix

\section{Emission and absorption line data}

In this appendix we give the emission and absorption line measurements
from the optical spectra as described in Section \ref{optspec}.

\begin{table*}
\footnotesize
\begin{center}
\begin{tabular}{llccrrrrl}
\hline\hline
\mc{1}{l}{Name} &\mc{1}{l}{line} &\mc{1}{c}{$\lambda_{\rm rest}$} &\mc{1}{c}{$\lambda_{\rm obs}$} &\mc{1}{c}{FWHM}           &\mc{1}{c}{flux} &\mc{1}{c}{snr}   &\mc{1}{c}{W$_{\lambda}$}\\  
\mc{1}{c}{} &\mc{1}{c}{ }    &\mc{1}{c}{(\AA)}  &\mc{1}{c}{(\AA)} &\mc{1}{c}{(\AA)}  &\mc{1}{c}{(W m$^{-2}$)} &\mc{1}{c}{$\sigma$} &\mc{1}{c}{(\AA)} \\
\hline\hline
{\bf 5C6.19}           &  [OII]         & 3727  & 6706 $\pm$ \f 1 &  21 $\pm$ \f 1 & 1.7e-19  &  3.2  &  78 $\pm$ \f 18   &  \\
$z=0.799 \pm 0.002$    &  CaII${_a}$    & 3934  & 7088 $\pm$ \f 7 & --\f\f\f\f--   & ---\f\f  &  2.0  &  --\f\f\f\f\f--   &  \\ 
$D(4000)=1.86 \pm 0.20$&  CaII${_a}$    & 3969  & 7128 $\pm$ \f 4 & --\f\f\f\f--   & ---\f\f  &  2.2  &  --\f\f\f\f\f--   &  \\ \hline 
{\bf 5C6.24}           &  CIII]         & 1909  & 3957 $\pm$ \f 2 &  37 $\pm$ \f 4 & 1.4e-19  &  1.1  &  99 $\pm$ \f 30   &  \\
$z=1.073\pm0.002$     &  CII]          & 2326  & 4822 $\pm$ \f 1 &  22 $\pm$ \f 1 & 1.3e-19  &  6    & 142 $\pm$ \f 24   &  \\
                       &  [OII]         & 3727  & 7721 $\pm$ \f 1 &  26 $\pm$ \f 3 & 5.2e-19  &  8    & 146 $\pm$ \f 20   &  \\
                       &  [NeIII]       & 3869  & 8012 $\pm$ \f 1 &  29 $\pm$ \f 4 & 8.5e-20  &  3.2  &  65 $\pm$ \f 18   &  \\ \hline
{\bf 5C6.25}           &  [NeIV]        & 2424  & 4139 $\pm$ \f 1 &  20 $\pm$ \f 2 & 5.4e-20  &  2.8  &  25 $\pm$ \f \f 9 &  \\ 
$z=0.706  \pm 0.002$   &  [NeV]         & 3426  & 5850 $\pm$ \f 1 &  19 $\pm$ \f 3 & 8.5e-20  &  2.5  &  51 $\pm$ \f 13   &  \\
$D(4000)=1.36 \pm 0.17$&  [OII]         & 3727  & 6362 $\pm$ \f 1 &  20 $\pm$ \f 1 & 8.0e-19  &  12   & 113 $\pm$ \f 15   &  \\
                       &  [NeIII]       & 3869  & 6600 $\pm$ \f 1 &  19 $\pm$ \f 3 & 1.8e-19  &  3.6  &  42 $\pm$ \f \f 9 &  \\  
                       &  H$\gamma$     & 4340  & 7407 $\pm$ \f 1 &  37 $\pm$ 11   & 1.5e-19  &  2.8  &  32 $\pm$ \f \f 8 &  \\ \hline
{\bf 5C6.29}           &  [OII]         & 3727  & 6409 $\pm$ \f 2 &  35 $\pm$ \f 4 & 3.8e-19  &  3.0  &  39 $\pm$ \f 11   &  \\
$z= 0.720\pm0.002$     &                &       &                 &                &          &       &                   &  \\
$D(4000)=1.73 \pm 0.17$&                &       &                 &                &          &       &                   &  \\ \hline
{\bf 5C6.43}           &  [OII]         & 3727  & 6635 $\pm$ \f 2 &  29 $\pm$ \f 3 & 3.3e-20  &  1.3  &  52 $\pm$ \f \f 8 &  \\
$z= 0.775 \pm  0.005$  &                &       &                 &                &          &       &                   &  \\
$D(4000)=2.43 \pm 0.40$&										                  &  \\ \hline
{\bf 5C6.63}           &  [OII]         & 3727  & 5451 $\pm$ \f 1 &  26 $\pm$ \f 2 & 8.0e-20  &  2.2  &  21 $\pm$ \f \f 8 &  \\ 
$z=0.465 \pm 0.003$    &  CaII${_a}$    & 3934  & 5798 $\pm$ \f 4 & --\f\f\f\f--   & ---\f\f  &  2.4  &  --\f\f\f\f\f--   &  \\
$D(4000)=1.91 \pm 0.09$&  CaII${_a}$    & 3969  & 5815 $\pm$ \f 2 & --\f\f\f\f--   & ---\f\f  &  2.0  &  --\f\f\f\f\f--   &  \\
	               &  [OIII]        & 5007  & 7347 $\pm$ \f 1 &  18 $\pm$ \f 1 & 1.4e-19  &  1.6  &   6 $\pm$ \f \f 2 &  \\ \hline
{\bf 5C6.75}           &  [OII]         & 3727  & 6617 $\pm$ \f 1 &  26 $\pm$ \f 1 & 1.9e-19  &  2.5  &  --\f\f\f\f\f--   &  \\
$z= 0.775 \pm  0.002$  &   	             								                  &  \\
$D(4000)=3.35 \pm 0.39$&  							                                          &  \\ \hline
{\bf 5C6.78}           &  [OII]         & 3727  & 4711 $\pm$ \f 2 &  32 $\pm$ \f 4 & 8.1e-19  &  8    &  38 $\pm$ \f 10   &  \\
$z= 0.263 \pm 0.002$   &  [NeIII]       & 3869  & 4881 $\pm$ \f 1 &  26 $\pm$ \f 2 & 2.5e-19  &  3.3  &  13 $\pm$ \f \f 3 &  \\ 
$D(4000)=1.98 \pm 0.05$&  H$\beta$      & 4861  & 6139 $\pm$ \f 1 &  16 $\pm$ \f 1 & 3.0e-19  &  3.3  &   5 $\pm$ \f \f 1 &  \\
	               &  [OIII]        & 4959  & 6261 $\pm$ \f 1 &  16 $\pm$ \f 2 & 1.4e-18  &  9    &  27 $\pm$ \f \f 3 &  \\
	               &  [OIII]        & 5007  & 6321 $\pm$ \f 1 &  13 $\pm$ \f 1 & 3.1e-18  & 19    &  64 $\pm$ \f 10   &  \\ \hline
{\bf 5C6.201}          &  CaII${_a}$    & 3969  & 6360 $\pm$ \f 5 & --\f\f\f\f--   & ---\f\f  &  2.7  &  --\f\f\f\f\f--   &  \\       
$z= 0.595 \pm 0.005$   &  [OIII]        & 4959  & 7939 $\pm$ \f 2 &  27 $\pm$ \f 3 & 1.0e-19  &  1.1  &  25 $\pm$ \f 10   &  \\       
$D(4000)=1.72 \pm 0.09$&  [OIII]        & 5007  & 7987 $\pm$ \f 3 &  26 $\pm$ \f 4 & 1.3e-19  &  1.5  &  30 $\pm$ \f 10   &  \\ \hline
{\bf 5C6.214}          &  [OII]         & 3727  & 5944 $\pm$ \f 1 &  17 $\pm$ \f 1 & 8.3e-20  &  2.3  &  39 $\pm$ \f \f 6 &  \\
$z= 0.595 \pm 0.002$   &  CaII${_a}$    & 3934  & 6277 $\pm$ \f 2 & --\f\f\f\f--   & ---\f\f  &  1.3  &  --\f\f\f\f\f--   &  \\
$D(4000)=2.06 \pm 0.12$&  CaII${_a}$    & 3969  & 6334 $\pm$ \f 3 & --\f\f\f\f--   & ---\f\f  &  2.7  &  --\f\f\f\f\f--   &  \\
	               &  H$\beta{_a}$  & 4861  & 7760 $\pm$ \f 3 & --\f\f\f\f--   & ---\f\f  &  2.2  &  --\f\f\f\f\f--   &  \\
 	               &  [OIII]        & 4959  & 7914 $\pm$ \f 1 &  19 $\pm$ \f 2 & 1.4e-19  &  1.9  &  19 $\pm$ \f \f 3 &  \\  
	               &  [OIII]        & 5007  & 7982 $\pm$ \f 2 &  25 $\pm$ \f 2 & 2.1e-19  &  2.7  &  30 $\pm$ \f \f 5 &  \\ \hline
{\bf 5C6.217}          &  HeII          & 1640  & 3960 $\pm$ \f 1 &  15 $\pm$ \f 1 & 1.7e-19  &  2.4  &  74 $\pm$ \f 30   &  \\
$z=1.410 \pm 0.003$    &  CIII]         & 1909  & 4599 $\pm$ \f 2 &  19 $\pm$ \f 1 & 6.5e-20  &  2.0  &  41 $\pm$ \f \f 7 &  \\
	               &  [NeIV]        & 2424  & 5831 $\pm$ \f 1 &  21 $\pm$ \f 1 & 7.6e-20  &  2.8  &  52 $\pm$ \f 14   &  \\  
	               &  [NeV]         & 3346  & 8060 $\pm$ \f 4 &  41 $\pm$ \f 1 & 1.1e-19  &  1.6  &  72 $\pm$ \f 14   &  \\
	               &  [NeV]         & 3426  & 8254 $\pm$ \f 1 &  28 $\pm$ \f 2 & 1.5e-19  &  3.2  &  97 $\pm$ \f 24   &  \\ \hline
{\bf 5C6.233}          &  [OII]         & 3727  & 5816 $\pm$ \f 1 &  25 $\pm$ \f 2 & 1.5e-19  &  5    &  86 $\pm$ \f 13   &  \\
$z=0.560\pm0.002$      &  [NeIII]       & 3869  & 6029 $\pm$ \f 1 &  24 $\pm$ \f 1 & 8.5e-20  &  3.1  &  39 $\pm$ \f \f 7 &  \\
$D(4000)=1.93 \pm 0.34$&                                                                                                  &  \\ \hline
{\bf 5C6.239}          &  CII]          & 2326  & 4217 $\pm$ \f 1 &  33 $\pm$ \f 2 & 8.9e-19  &  1.3  &  --\f\f\f\f\f--   &  \\
$z=0.805\pm0.002$      &  [NeIV]        & 2424  & 4369 $\pm$ \f 2 &  28 $\pm$ \f 1 & 9.1e-19  &  3.2  &  --\f\f\f\f\f--   &  \\
	               &  MgII          & 2798  & 5048 $\pm$ \f 1 &  26 $\pm$ \f 3 & 9.8e-20  &  1.8  &  --\f\f\f\f\f--   &  \\
	               &  [NeV]         & 3346  & 6041 $\pm$ \f 1 &  41 $\pm$ \f 4 & 7.1e-20  &  2.0  &  70 $\pm$ \f 22   &  \\ 
	               &  [NeV]         & 3426  & 6184 $\pm$ \f 1 &  39 $\pm$ \f 2 & 1.4e-19  &  3.3  &  64 $\pm$ \f 12   &  \\ 
	               &  [OII]         & 3727  & 6728 $\pm$ \f 1 &  21 $\pm$ \f 1 & 4.0e-19  &  18   & 121 $\pm$ \f 11   &  \\
	               &  [NeIII]       & 3869  & 6985 $\pm$ \f 1 &  32 $\pm$ \f 3 & 1.1e-19  &  5    &  95 $\pm$ \f 30   &  \\ \hline
{\bf 5C6.258}          &  [OII]         & 3727  & 6528 $\pm$ \f 1 &  21 $\pm$ \f 1 & 1.3e-19  &  6    &  --\f\f\f\f\f--   &  \\ 
$z=0.752 \pm 0.002$    &												  &  \\
$D(4000)=1.48 \pm 0.37$&                 									          &  \\ 

\hline\hline         
\end{tabular}
\end{center}              

{\caption[Table of observations]{\label{tab:emli} (cont.) Table
summarizing the emission and absorption lines present in the optical
spectra of the 7C Redshift Survey radio galaxies. The subscript $a$
denotes that the observed line is seen in absorption, not
emission. Values of the 4000 \AA\ break parameter $D(4000)$ are given
for all $0.2<z<0.8$ radio galaxies (see Section 5.2). }}

\normalsize
\end{table*}

\addtocounter{table}{-1}

\begin{table*}
\footnotesize
\begin{center}
\begin{tabular}{llccrrrrl}
\hline\hline
\mc{1}{c}{name} &\mc{1}{l}{line} &\mc{1}{c}{$\lambda_{\rm rest}$} &\mc{1}{c}{$\lambda_{\rm obs}$} &\mc{1}{c}{FWHM}           &\mc{1}{c}{flux} &\mc{1}{c}{snr}   &\mc{1}{c}{W$_{\lambda}$}\\  
\mc{1}{c}{} &\mc{1}{c}{ }    &\mc{1}{c}{(\AA)}  &\mc{1}{c}{(\AA)} &\mc{1}{c}{(\AA)}  &\mc{1}{c}{(W m$^{-2}$)} &\mc{1}{c}{$\sigma$} &\mc{1}{c}{(\AA)} \\
\hline\hline
{\bf 5C6.267}          &  [OII]         & 3727  & 5053 $\pm$ \f 2 &  28 $\pm$ \f 2 & 1.2e-18  &  3.1  &  49 $\pm$ \f \f 7 &  \\ 
$z=0.357\pm 0.002$     &  CaII${_a}$    & 3934  & 5331 $\pm$ \f 1 & --\f\f\f\f--   & ---\f\f  &  2.2  &  --\f\f\f\f\f--   &  \\ 
$D(4000)=1.93 \pm 0.12$&  CaII${_a}$    & 3969  & 5370 $\pm$ \f 3 & --\f\f\f\f--   & ---\f\f  &  1.4  &  --\f\f\f\f\f--   &  \\ 
	               &  G${_a}$       & 4304  & 5832 $\pm$ \f 4 & --\f\f\f\f--   & ---\f\f  &  2.1  &  --\f\f\f\f\f--   &  \\
   	               &  H$\beta$${_a}$& 4861  & 6607 $\pm$ \f 1 & --\f\f\f\f--   & ---\f\f  &  3.1  &  --\f\f\f\f\f--   &  \\
	               &  NaI${_a}$     & 5893  & 7983 $\pm$ \f 1 & --\f\f\f\f--   & ---\f\f  &  3.5  &  --\f\f\f\f\f--   &  \\ \hline
{\bf 5C6.279}          &  [NeV]         & 3426  & 5067 $\pm$ \f 8 &  60 $\pm$   15 & 1.5e-19  &  3.1  &  89 $\pm$ \f 11   &  \\
$z=0.473 \pm 0.003$    &  [OII]	        & 3727  & 5504 $\pm$ \f 1 &  22 $\pm$ \f 2 & 5.9e-19  &  4    &  --\f\f\f\f\f--   &  \\
$D(4000)=2.36 \pm 0.28$&  H$\beta$      & 4861  & 7151 $\pm$ \f 1 &  21 $\pm$ \f 1 & 9.7e-20  &  2.5  &  10 $\pm$ \f \f 3 &  \\
	               &  [OIII]        & 4959  & 7300 $\pm$ \f 1 &  20 $\pm$ \f 1 & 3.2e-19  &    9  &  19 $\pm$ \f \f 1 &  \\ 
	               &  [OIII]        & 5007  & 7371 $\pm$ \f 1 &  19 $\pm$ \f 1 & 7.1e-19  &   21  &  44 $\pm$ \f \f 5 &  \\ \hline
{\bf 7C0221+3417}      &  CII]          & 2326  & 4305 $\pm$ \f 2 &  13 $\pm$ \f 1 & 3.1e-20  &  1.5  &  --\f\f\f\f\f--   &  \\
$z=0.852 \pm 0.002$    &  [OII]         & 3727  & 6900 $\pm$ \f 1 &  23 $\pm$ \f 1 & 1.2e-19  &  4    &  --\f\f\f\f\f--   &  \\
	               &  [NeIII]       & 3869  & 7165 $\pm$ \f 1 &  29 $\pm$ \f 1 & 3.7e-20  &  1.2  &  --\f\f\f\f\f--   &  \\ \hline
{\bf 5C6.292}          &  HeII          & 1640  & 3674 $\pm$ \f 2 &  15 $\pm$ \f 2 & 1.3e-19  &  2.6  &  --\f\f\f\f\f--   &  \\
$z=1.245 \pm 0.004$    &  CIII]         & 1909  & 4275 $\pm$ \f 2 &  34 $\pm$ \f 2 & 3.2e-19  &    5  &  --\f\f\f\f\f--   &  \\
	               &  [OII]         & 3727  & 8379 $\pm$ \f 2 &  29 $\pm$ \f 1 & 9.0e-19  &    8  &  --\f\f\f\f\f--   &  \\ \hline
{\bf 5C7.7}            &  [OII]	        & 3727  & 5348 $\pm$ \f 9 &  32 $\pm$ \f 2 & 2.0e-19  &  2.3  &  15 $\pm$ \f \f 3 &  \\
$z=0.435 \pm 0.002$    &  CaII${_a}$    & 3934  & 5645 $\pm$ \f 1 & --\f\f\f\f--   & ---\f\f  &  2.6  &  --\f\f\f\f\f--   &  \\ 
$D(4000)=1.63 \pm 0.07$&  CaII${_a}$    & 3969  & 5699 $\pm$ \f 1 & --\f\f\f\f--   & ---\f\f  &  2.8  &  --\f\f\f\f\f--   &  \\ 
	               &  G${_a}$       & 4304  & 6178 $\pm$ \f 1 & --\f\f\f\f--   & ---\f\f  &  1.0  &  --\f\f\f\f\f--   &  \\
	               &  H$\beta$      & 4861  & 6989 $\pm$ \f 5 &  38 $\pm$ \f 8 & 1.1e-19  &  0.8  &   7 $\pm$ \f \f 2 &  \\ 
	               &  [OIII]        & 4959  & 7114 $\pm$ \f 3 &  48 $\pm$ \f 2 & 3.3e-19  &  2.7  &  18 $\pm$ \f \f 3 &  \\
	               &  [OIII]        & 5007  & 7183 $\pm$ \f 2 &  32 $\pm$ \f 2 & 6.8e-19  &    6  &  33 $\pm$ \f \f 6 &  \\
	               &  MgI${_a}$     & 5175  & 7429 $\pm$ \f 2 & --\f\f\f\f--   & ---\f\f  &  1.3  &  --\f\f\f\f\f--   &  \\ \hline 
{\bf 5C7.8}            &  [OII]	        & 3727  & 6236 $\pm$ \f 1 &  19 $\pm$ \f 1 & 3.8e-19  &  2.4  &  83 $\pm$ \f 13   &  \\
$z=  0.673 \pm0.002$   &  CaII${_a}$    & 3934  & 6585 $\pm$ \f 2 & --\f\f\f\f--   & ---\f\f  &  1.4  &  --\f\f\f\f\f--   &  \\ 
$D(4000)=2.23 \pm 0.12$&  CaII${_a}$    & 3969  & 6646 $\pm$ \f 4 & --\f\f\f\f--   & ---\f\f  &  1.3  &  --\f\f\f\f\f--   &  \\ \hline
{\bf 5C7.9}            &  CaII${_a}$    & 3934  & 4852 $\pm$ \f 2 & --\f\f\f\f--   & ---\f\f  &  1.4  &  --\f\f\f\f\f--   &  \\
 $z=0.233  \pm 0.002$  &  G${_a}$       & 4304  & 5308 $\pm$ \f 2 & --\f\f\f\f--   & ---\f\f  &    5  &  --\f\f\f\f\f--   &  \\
$D(4000)=2.62 \pm 0.25$&  [OIII]        & 4959  & 6116 $\pm$ \f 1 &  18 $\pm$ \f 1 & 5.2e-18  &  8    &  27 $\pm$ \f \f 3 &  \\
	               &  [OIII]        & 5007  & 6174 $\pm$ \f 1 &  16 $\pm$ \f 1 & 1.3e-17  & 24    &  62 $\pm$ \f \f 9 &  \\ \hline
{\bf 5C7.10}           &  Ly$\alpha$    & 1216  & 3874 $\pm$ \f 1 &  28 $\pm$ \f 1 & 2.7e-18  & 12    &  --\f\f\f\f\f--   &  \\ 
$z=2.185\pm0.007$      &  CIV	        & 1549  & 4935 $\pm$ \f 1 &  41 $\pm$ \f 3 & 2.0e-19  &  2.9  &  --\f\f\f\f\f--   &  \\ 
	               &  HeII          & 1640  & 5206 $\pm$ \f 2 &  12 $\pm$ \f 1 & 1.9e-19  &  3.1  &  --\f\f\f\f\f--   &  \\
	               &  CIII]	        & 1909  & 6049 $\pm$ \f 1 &  25 $\pm$ \f 2 & 1.5e-19  &  4    &  --\f\f\f\f\f--   &  \\ \hline
{\bf 5C7.15}           &  Ly$\alpha$    & 1216  & 4175 $\pm$ \f 1 &  24 $\pm$ \f 1 & 4.2e-19  &  6    &  --\f\f\f\f\f--   &  \\
 $z=2.433 \pm 0.002$   &												  &  \\ \hline
{\bf 5C7.23}           &  CIII]	        & 1909  & 3988 $\pm$ \f 1 &  19 $\pm$ \f 4 & 4.7e-20  &  0.9  &  72 $\pm$ \f 13   &  \\ 
 $z=1.098 \pm  0.005$  &  MgII          & 2798  & 5857 $\pm$ \f 4 &  40 $\pm$ \f 1 & 5.0e-20  &  2.5  &  --\f\f\f\f\f--   &  \\ 
	               &  [OII]         & 3727  & 7820 $\pm$ \f 2 &  18 $\pm$ \f 1 & 2.0e-19  &  4    &  --\f\f\f\f\f--   &  \\ \hline
{\bf 5C7.25}           &  [OII]	        & 3727  & 6225 $\pm$ \f 3 &  27 $\pm$ \f 2 & 2.4e-19  &  6    &  104 $\pm$ \f 30  &  \\ 
$z=0.671 \pm 0.003$    &  CaII${_a}$    & 3934  & 6564 $\pm$ \f 1 &  --\f\f\f\f--  & ---\f\f  &  2.1  &  --\f\f\f\f\f--   &  \\  
$D(4000)=1.51 \pm 0.18$&  CaII${_a}$    & 3969  & 6624 $\pm$ \f 1 &  --\f\f\f\f--  & ---\f\f  &  2.0  &  --\f\f\f\f\f--   &  \\
	               &  G${_a}$       & 4304  & 7189 $\pm$ \f 2 &  --\f\f\f\f--  & ---\f\f  &  3.1  &  --\f\f\f\f\f--   &  \\ \hline
{\bf 5C7.57}           &  CIV	        & 1549  & 4067 $\pm$ \f 2 &  38 $\pm$ \f 2 & 1.7e-19  &  4    &  --\f\f\f\f\f--   &  \\
$z=1.622 \pm 0.003$    &  CIII]	        & 1909  & 5003 $\pm$ \f 2 &  25 $\pm$ \f 1 & 6.1e-20  &  2.8  &  --\f\f\f\f\f--   &  \\
	               &  MgII 	        & 2798  & 7356 $\pm$ \f 2 &  35 $\pm$ \f 2 & 4.2e-20  &  2.6  &  --\f\f\f\f\f--   &  \\ \hline
{\bf 5C7.78}           &  CIII]	        & 1909  & 4107 $\pm$ \f 3 &  33 $\pm$ \f 7 & 8.0e-20  &  1.3  &  --\f\f\f\f\f--   &  \\
$z=1.151 \pm 0.002$    &  [NeV]         & 3426  & 7340 $\pm$ \f 1 &  18 $\pm$ \f 1 & 5.9e-20  &  1.0  &  --\f\f\f\f\f--   &  \\
	               &  [OII]         & 3727  & 8032 $\pm$ \f 2 &  38 $\pm$ \f 3 & 3.7e-19  &  5    &  --\f\f\f\f\f--   &  \\
                       &  [NeIII]       & 3869  & 8325 $\pm$ \f 3 &  29 $\pm$ \f 4 & 3.1e-19  &  4    &  --\f\f\f\f\f--   &  \\ \hline 
{\bf 5C7.79}           &  [OII]	        & 3727  & 5994 $\pm$ \f 1 &  20 $\pm$ \f 5 & 8.9e-20  &  3.4  &  68 $\pm$ \f 19   &  \\
$z=0.608 \pm 0.003$    &  CaII${_a}$    & 3934  & 6315 $\pm$ \f 1 &  --\f\f\f\f--  & ---\f\f  &  2.6  &  --\f\f\f\f\f--   &  \\
$D(4000)=2.10 \pm 0.21$&  CaII${_a}$    & 3969  & 6379 $\pm$ \f 1 &  --\f\f\f\f--  & ---\f\f  &  1.8  &  --\f\f\f\f\f--   &  \\
	               &  G${_a}$       & 4304  & 6921 $\pm$ \f 5 &  --\f\f\f\f--  & ---\f\f  &  2.5  &  --\f\f\f\f\f--   &  \\ \hline 
{\bf 5C7.82}           &  [OII]	        & 3727  & 7147 $\pm$ \f 1 &  32 $\pm$ \f 5 & 1.5e-19  &  3.5  &  --\f\f\f\f\f--   &  \\
$z=0.918 \pm 0.002$    & 										                  &  \\ 
\hline\hline         
\end{tabular}
\end{center}              
{\caption[Table of observations]{cont.}}
\normalsize
\end{table*}

\addtocounter{table}{-1}

\begin{table*}
\footnotesize
\begin{center}
\begin{tabular}{llccrrrrl}
\hline\hline
\mc{1}{c}{name} &\mc{1}{l}{line} &\mc{1}{c}{$\lambda_{\rm rest}$} &\mc{1}{c}{$\lambda_{\rm obs}$} &\mc{1}{c}{FWHM}           &\mc{1}{c}{flux} &\mc{1}{c}{snr}   &\mc{1}{c}{W$_{\lambda}$}\\  
\mc{1}{c}{} &\mc{1}{c}{ }    &\mc{1}{c}{(\AA)}  &\mc{1}{c}{(\AA)} &\mc{1}{c}{(\AA)}  &\mc{1}{c}{(W m$^{-2}$)} &\mc{1}{c}{$\sigma$} &\mc{1}{c}{(\AA)} \\
\hline\hline
{\bf 5C7.106}          &  CaII${_a}$    & 3934  & 4975 $\pm$ \f 1 &  --\f\f\f\f--  & ---\f\f  &    5  &  --\f\f\f\f\f--   &  \\
$z=0.264 \pm 0.002$    &  CaII${_a}$    & 3969  & 5012 $\pm$ \f 1 &  --\f\f\f\f--  & ---\f\f  &  1.1  &  --\f\f\f\f\f--   &  \\
$D(4000)=2.05 \pm 0.06$&  G${_a}$       & 4304  & 5439 $\pm$ \f 1 &  --\f\f\f\f--  & ---\f\f  &  1.4  &  --\f\f\f\f\f--   &  \\
                       &  MgI${_a}$     & 5175  & 6541 $\pm$ \f 1 &  --\f\f\f\f--  & ---\f\f  &  2.7  &  --\f\f\f\f\f--   &  \\ \hline
{\bf 5C7.111}          &  [OII]	        & 3727  & 6068 $\pm$ \f 1 &  17 $\pm$ \f 1 & 2.4e-19  &    7  &  86 $\pm$ \f 14   &  \\ 
$z=0.628 \pm 0.002$    &  [OIII]        & 4959  & 8070 $\pm$ \f 1 &  16 $\pm$ \f 1 & 1.7e-19  &    6  &  20 $\pm$ \f \f 3 &  \\  
$D(4000)=2.00 \pm 0.11$&  [OIII]        & 5007  & 8149 $\pm$ \f 1 &  19 $\pm$ \f 1 & 5.0e-19  &   11  &  58 $\pm$ \f \f 6 &  \\ \hline 
{\bf 5C7.125}          &  [OII]	        & 3727  & 6715 $\pm$ \f 1 &  16 $\pm$ \f 1 & 1.2e-19  &  6    &  151 $\pm$ \f 35  &  \\
$z=0.801 \pm 0.002$    &					  						          &  \\ \hline
{\bf 5C7.145}          &  CaII${_a}$    & 3934  & 5283 $\pm$ \f 2 &  --\f\f\f\f--  & ---\f\f  &  2.6  &  --\f\f\f\f\f--   &  \\
$z=0.343 \pm 0.002$    &  CaII${_a}$    & 3969  & 5329 $\pm$ \f 1 &  --\f\f\f\f--  & ---\f\f  &  2.4  &  --\f\f\f\f\f--   &  \\
$D(4000)=2.31 \pm 0.08$&  G${_a}$       & 4304  & 5780 $\pm$ \f 1 &  --\f\f\f\f--  & ---\f\f  &  1.7  &  --\f\f\f\f\f--   &  \\ 
	               &  H$\beta$${_a}$& 4861  & 6533 $\pm$ \f 2 &  --\f\f\f\f--  & ---\f\f  &  2.3  &  --\f\f\f\f\f--   &  \\
	               &  MgI${_a}$     & 5175  & 6949 $\pm$ \f 4 &  --\f\f\f\f--  & ---\f\f  &  3.3  &  --\f\f\f\f\f--   &  \\
	               &  NaI${_a}$     & 5893  & 7913 $\pm$ \f 1 &  --\f\f\f\f--  & ---\f\f  &  3.1  &  --\f\f\f\f\f--   &  \\ \hline
{\bf 5C7.170}          &  CaII${_a}$    & 3934  & 4987 $\pm$ \f 3 &  --\f\f\f\f--  & ---\f\f  &  2.7  &  --\f\f\f\f\f--   &  \\ 
$z=0.268 \pm 0.002$    &  CaII${_a}$    & 3969  & 5030 $\pm$ \f 1 &  --\f\f\f\f--  & ---\f\f  &  2.7  &  --\f\f\f\f\f--   &  \\
$D(4000)=2.48 \pm 0.11$&  G${_a}$       & 4304  & 5460 $\pm$ \f 2 &  --\f\f\f\f--  & ---\f\f  &  3.2  &  --\f\f\f\f\f--   &  \\
	               &  MgI${_a}$     & 5175  & 6565 $\pm$ \f 2 &  --\f\f\f\f--  & ---\f\f  &  3.5  &  --\f\f\f\f\f--   &  \\
	               &  NaI${_a}$     & 5893  & 7477 $\pm$ \f 1 &  --\f\f\f\f--  & ---\f\f  &  3.0  &  --\f\f\f\f\f--   &  \\ \hline 
{\bf 5C7.178}          &  [OII]	        & 3727  & 4642 $\pm$ \f 1 &  14 $\pm$ \f 1 & 1.0e-19  &  4    &  18 $\pm$ \f \f 2 &  \\ 
$z=0.246 \pm 0.002$    &  [NeIII]       & 3869  & 4805 $\pm$ \f 1 &  20 $\pm$ \f 2 & 1.2e-19  &  2.5  &  23 $\pm$ \f \f 3 &  \\  
$D(4000)=1.74 \pm 0.09$&  CaII${_a}$    & 3934  & 4898 $\pm$ \f 1 &  --\f\f\f\f--  & ---\f\f  &  1.3  &  --\f\f\f\f\f--   &  \\
	               &  CaII${_a}$    & 3969  & 4945 $\pm$ \f 1 &  --\f\f\f\f--  & ---\f\f  &  3.0  &  --\f\f\f\f\f--   &  \\
	               &  [OIII]        & 4959  & 6186 $\pm$ \f 3 &  25 $\pm$ \f 1 & 1.9e-20  &  1.4  &  5 $\pm$ \f \f 2  &  \\
	               &  [OIII]        & 5007  & 6236 $\pm$ \f 1 &  22 $\pm$ \f 5 & 2.5e-20  &  1.7  &  7 $\pm$ \f \f 2  &  \\
	               &  NaI${_a}$     & 5893  & 7341 $\pm$ \f 1 &  --\f\f\f\f--  & ---\f\f  &  2.4  &  --\f\f\f\f\f--   &  \\
	               &  H$\alpha$     & 6563  & 8175 $\pm$ \f 1 &  14 $\pm$ \f 1 & 2.5e-19  &   10  &  37 $\pm$ \f \f 4 &  \\ \hline
{\bf 5C7.205}          &  [OII]	        & 3727  & 6375 $\pm$ \f 2 &  24 $\pm$ \f 3 & 5.8e-19  &   10  &  157 $\pm$ \f 44  &  \\
$z=0.710 \pm 0.002$    &  [NeIII]       & 3869  & 6614 $\pm$ \f 2 &  21 $\pm$ \f 4 & 2.4e-19  &  3.4  &  --\f\f\f\f\f--   &  \\
$D(4000)=2.09 \pm 0.19$&  H$\gamma$     & 4340  & 7429 $\pm$ \f 3 &  30 $\pm$ \f 1 & 2.3e-19  &  3.4  &  81 $\pm$ \f 30   &  \\ \hline
{\bf 5C7.223}          &  Ly$\alpha$    & 1216  & 3760 $\pm$ \f 3 &  26 $\pm$ \f 2 & 1.7e-18  &   16  &  --\f\f\f\f\f--   &  \\
$z=2.092 \pm  0.005$   &  CIV	        & 1549  & 4789 $\pm$ \f 1 &  15 $\pm$ \f 3 & 2.6e-20  &  2.8  &  --\f\f\f\f\f--   &  \\
	               &  HeII	        & 1640  & 5066 $\pm$ \f 3 &  27 $\pm$ \f 1 & 7.4e-20  &  2.9  &  --\f\f\f\f\f--   &  \\ \hline
{\bf 5C7.242}          &  CIII]	        & 1909  & 3814 $\pm$ \f 1 &   8 $\pm$ \f 1 & 3.2e-20  &  1.2  &  --\f\f\f\f\f--   &  \\
$z=0.992 \pm 0.006$    &  [OII]         & 3727  & 7427 $\pm$ \f 1 &  23 $\pm$ \f 1 & 9.0e-20  &  3.4  &  --\f\f\f\f\f--   &  \\ \hline
{\bf 5C7.269}          &  Ly$\alpha$    & 1216  & 3909 $\pm$ \f 1 &  21 $\pm$ \f 3 & 9.4e-19  &  9.2  &  --\f\f\f\f\f--   &  \\
$z=2.218 \pm 0.003$    &  											          &  \\ \hline
{\bf 7C0825+2446}      &  CaII${_a}$    & 3934  & 4272 $\pm$ \f 1 &  --\f\f\f\f--  & ---\f\f  &  2.7  &  --\f\f\f\f\f--   &  \\
$z=0.086 \pm 0.002$    &  CaII${_a}$    & 3969  & 4314 $\pm$ \f 2 &  --\f\f\f\f--  & ---\f\f  &  2.8  &  --\f\f\f\f\f--   &  \\
	               &  G${_a}$       & 4304  & 4667 $\pm$ \f 1 &  --\f\f\f\f--  & ---\f\f  &  3.8  &  --\f\f\f\f\f--   &  \\
	               &  H$\beta$${_a}$& 4861  & 5283 $\pm$ \f 1 &  --\f\f\f\f--  & ---\f\f  &  3.2  &  --\f\f\f\f\f--   &  \\
	               &  MgI${_a}$     & 5175  & 5622 $\pm$ \f 1 &  --\f\f\f\f--  & ---\f\f  &    5  &  --\f\f\f\f\f--   &  \\
	               &  NaI${_a}$     & 5893  & 6399 $\pm$ \f 1 &  --\f\f\f\f--  & ---\f\f  &    8  &  --\f\f\f\f\f--   &  \\ \hline
{\bf 7C0825+2443}      &  [OII]         & 3727  & 4628 $\pm$ \f 2 &  30 $\pm$ \f 2 & 9.9e-19  &  2.1  &  42 $\pm$ \f 11   &  \\
$z=0.243 \pm 0.002$    &  CaII${_a}$    & 3934  & 4893 $\pm$ \f 1 &  --\f\f\f\f--  & ---\f\f  &    5  &  --\f\f\f\f\f--   &  \\
$D(4000)=2.11 \pm 0.12$&  CaII${_a}$    & 3969  & 4931 $\pm$ \f 1 &  --\f\f\f\f--  & ---\f\f  &    4  &  --\f\f\f\f\f--   &  \\
	               &  G${_a}$	& 4304  & 5350 $\pm$ \f 2 &  --\f\f\f\f--  & ---\f\f  &  1.9  &  --\f\f\f\f\f--   &  \\
	               &  MgI${_a}$     & 5175  & 6433 $\pm$ \f 2 &  --\f\f\f\f--  & ---\f\f  &    4  &  --\f\f\f\f\f--   &  \\ 
	               &  NaI${_a}$     & 5893  & 7318 $\pm$ \f 1 &  --\f\f\f\f--  & ---\f\f  &  2.7  &  --\f\f\f\f\f--   &  \\ \hline
{\bf 5C7.271}          &  Ly$\alpha$    & 1216  & 3910 $\pm$ \f 2 &  49 $\pm$ \f 2 & 9.6e-20  &  3.8  &  --\f\f\f\f\f--   &  \\
$z=2.224 \pm 0.006$    &  CIV           & 1549  & 4981 $\pm$ \f 3 &  28 $\pm$ \f 2 & 3.2e-20  &  3.4  &  --\f\f\f\f\f--   &  \\
	               &  HeII	        & 1640  & 5305 $\pm$ \f 1 &  27 $\pm$ \f 5 & 3.2e-20  &  2.4  &  --\f\f\f\f\f--   &  \\
	               &  CII]	        & 2326  & 7498 $\pm$ \f 4 &  47 $\pm$ \f 1 & 3.5e-20  &  2.7  &  --\f\f\f\f\f--   &  \\
	               &  [NeIV]        & 2424  & 7810 $\pm$ \f 1 &  16 $\pm$ \f 1 & 2.1e-20  &  2.0  &  --\f\f\f\f\f--   &  \\
\hline\hline         
\end{tabular}
\end{center}              
{\caption[Table of observations]{cont.}}
\normalsize
\end{table*}

\end{document}